\documentclass[aps,prl,reprint,showpacs,superscriptaddress]{revtex4-1}
\usepackage[english]{babel}
\usepackage{amsmath,amssymb,bbm,graphicx,color,comment,txfonts,braket,mathtools}
\usepackage[bookmarks=true,colorlinks,citecolor=blue,urlcolor=blue]{hyperref}
\usepackage{dsfont}
\usepackage{tikz}

\usepackage{xr}
\begin{document}

\date{\today}

\title{Hydrodynamic stabilization of self-organized criticality in a driven Rydberg gas}
\author{K. Klocke}
\affiliation{Department of Physics and Institute for Quantum Information and Matter, California Institute of Technology, Pasadena, CA 91125, USA}
\affiliation{Department of Physics, University of California, Berkeley, California 94720, USA}
\author{T.~M. Wintermantel}
\affiliation{ISIS (UMR 7006), University of Strasbourg and CNRS, 67000 Strasbourg, France}
\affiliation{Physikalisches Institut, Universit\"at Heidelberg, 69120 Heidelberg, Germany}
\author{G. Lochead}
\affiliation{ISIS (UMR 7006), University of Strasbourg and CNRS, 67000 Strasbourg, France}
\author{S. Whitlock}
\affiliation{ISIS (UMR 7006), University of Strasbourg and CNRS, 67000 Strasbourg, France}
\author{M. Buchhold}
\affiliation{Department of Physics and Institute for Quantum Information and Matter, California Institute of Technology, Pasadena, CA 91125, USA}
\affiliation{Institut f\"ur Theoretische Physik, Universit\"at zu K\"oln, D-50937 Cologne, Germany}

\begin{abstract}
Signatures of self-organized criticality (SOC) have recently been observed in an ultracold atomic gas under continuous laser excitation to strongly-interacting Rydberg states~[S. Helmrich \textit{et al.}, Nature, \textbf{577}, 481--\,486 (2020)]. This creates a unique possibility to study this intriguing dynamical phenomenon, e.g., to probe its robustness and universality, under controlled experimental conditions. Here we examine the self-organizing dynamics of a driven ultracold gas and identify an unanticipated feedback mechanism, which is especially important for systems coupled to thermal baths. It sustains an extended critical region in the trap center for a notably long time via hydrodynamic transport of particles from the flanks of the cloud toward the center. This compensates the avalanche-induced atom loss and leads to a characteristic flat-top density profile, providing an additional experimental signature for SOC and minimizing effects of inhomogeneity on the SOC features.
\end{abstract}

\maketitle
{\it{Introduction.}}-- Many-body systems, may they be driven, open or excited by a sudden parameter quench, often evolve toward steady or transient metastable states which can be classified as far from thermal equilibrium~\cite{Janssen1981,Grassberger1983,Babadi2015,Tsatsos2016,Kadau2016,gillman2020,Munoz1999,Takeuchi2007,Lemoult2016,Sano2016,Nowak,Bak1987,buchhold2015}. Sometimes these systems feature attractors for the non-equilibrium dynamics that give rise to emergent scale invariant properties over a wide range of initial states or parameters~\cite{Berges2008,Kinouchi2006,Bertsch2004,Bornholdt2000,Nicklas2015}. One paradigmatic example is self-organized criticality (SOC), whereby a dissipative many-body system evolves toward a (non-equilibrium) critical state by an intrinsic feedback mechanism. Since its first introduction by Bak, Tang, and Wiesenfeld in 1987 ~\cite{Bak1987,Bak1988}, SOC has been intensively studied theoretically and associated with phenomena ranging from avalanches and earthquakes to solar flares and neuronal activity to name a few~\cite{Altshuler2004,Field1995,Swingle,Aschwanden2016,Turcotte1999}.

The range of phenomena found to exhibit SOC-like characteristics is at odds however with the relatively stringent conditions that are expected to lead to SOC in theory ~\cite{pruessner2012self}. For example, the typical requirements of a large separation of timescales between slow dissipation and fast, conservative bulk dynamics will never be perfectly satisfied in practice~\cite{Bonachela2009}. This has lead to the notion of self-organized quasi-criticality (SOqC) where the system hovers around criticality with large excursions into the sub- and super-critical phases~\cite{Bonachela2009,Bonachela2010}. Nonetheless, key signatures of self-organized criticality including scale invariance of the stationary density and power-law distributed excitation avalanches were recently observed in the driven-dissipative dynamics of atomic Rydberg gases~\cite{Helmrich2019} (see also related experiments in driven thermal gases~\cite{SOCAdams}). These experiments, however, lacked an obvious refilling mechanism necessary to bring the system out of the sub-critical absorbing phase. This therefore raises important questions about how signatures of the SOC state persist for relatively long times and whether possible universal characteristics of the SOC state~\cite{pruessner2012self} can be extracted from experiments in a transient regime.

\begin{figure}[!ht]
    \includegraphics[width=0.99\linewidth]{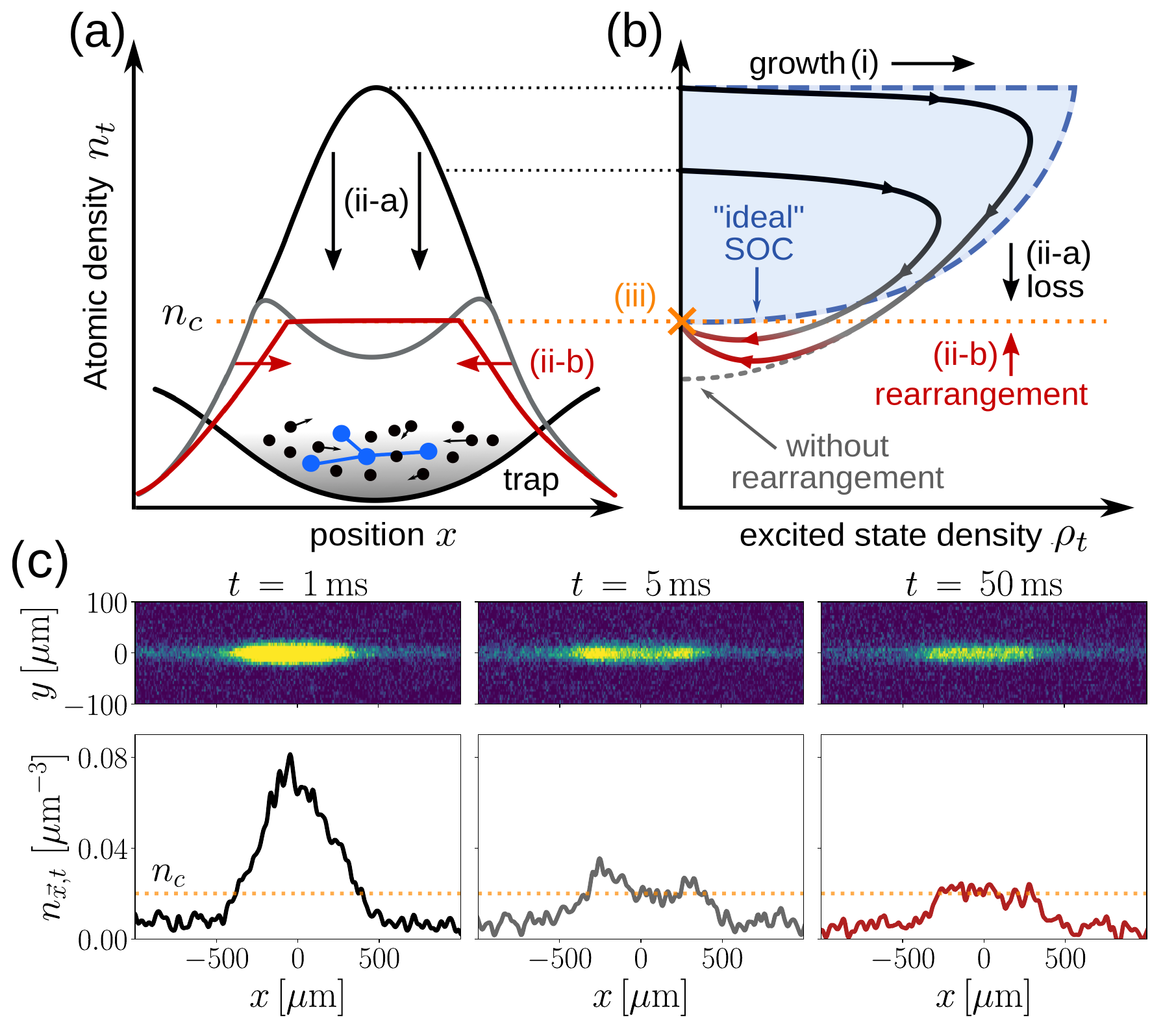}
    \caption{Mechanisms for self-organized criticality in an ultracold atomic gas. (a) A trapped atomic gas with inhomogeneous density distribution is continuously driven by an off-resonant excitation laser to highly excited Rydberg states (blue disks). (b) Trajectory of the atom density $n_t$ and the excitation density $\rho_t$ driven by facilitated excitation, decay and hydrodynamic motion. Starting from the supercritical phase $n_{t=0}>n_c$ the system undergoes: (i) rapid growth of Rydberg density; (ii-a) self-organization from the active phase toward the critical point due to gradual depletion of particles (caused by loss from the Rydberg state); (ii-b) refilling of the central density from the sub-critical phase by atomic rearrangement (thermal motion) from the wings to the trap center; (iii) stabilization close to the critical point for an extended period of time.  (c) upper panels: experimental absorption image ($n_{\vec{x},t}$ integrated over $z$) at different times. lower panels: Reconstructed three-dimensional atom density $n_{\vec{x},t}$ at $y=z=0$ showing the appearance of a flat-top profile coinciding with the SOC critical density.}
\label{fig:soc_mechanism}
\end{figure}

Here we experimentally and theoretically demonstrate that the mechanisms leading to the SOC state are remarkably robust. We show that the slow motion of the particles provides an additional feedback mechanism which stabilizes the system close to the critical state over an extensive period of time. This is evidenced by the experimental observation of a stable flat-top profile in the atomic gas, where the wings of the distribution act as particle reservoirs that compensate particle loss in the trap center (Fig.~\ref{fig:soc_mechanism}). To explain this result we develop a hydrodynamic Langevin equation which describes the competition between thermalization of the gas (in the motional degrees of freedom) and the driven-dissipative excitation dynamics leading to SOC. This allows the cloud to adapt by slowly refilling sub-critical regions back to a critical state, which plays a similar role to plasticity in biological neural networks~\cite{Levina2007,Zierenberg}.

{\it Self-organization mechanism.}-- We consider a spatially \mbox{inhomogeneous} gas of ultracold atoms held in a harmonic optical potential produced by a focused far-off-resonant laser beam~\cite{Helmrich2019} (depicted in Fig.~\ref{fig:soc_mechanism}a). The atoms are continuously driven by a detuned laser field, which creates rare and isolated Rydberg excitations at random positions in the gas. Once an excitation is present it will either spontaneously decay (which is often accompanied by loss from the trap), or it can trigger secondary excitations  through a process called Rydberg facilitation~\cite{Schempp2014,Malossi2014,Goldschmidt2016,Helmrich2018}. This occurs at a characteristic distance $r_\text{fac}\approx 4.5\, \mu$m (for the present experiments) where the laser detuning is compensated by the van der Waals interaction between Rydberg pair states \footnote{The facilitation radius can be approximated as $r_\text{fac} = \left(C_6 / \Delta\right)^{1/6}$, where $\Delta$ is the detuning and $C_6$ is the van der Waals coefficient.}.
The self-organizing dynamics are driven by the competition between facilitated excitation (with rate proportional to $\kappa n_{\vec{x},t}$, where $\kappa$ is the microscopic facilitation rate and $n_{\vec{x},t}$ is the local density of ground-state atoms) and density independent spontaneous decay or loss of the excited atoms with rate $\Gamma$. These two processes compete to produce rich collective dynamics~\cite{Lee2012,Lesanovsky2013,Marcuzzi2015,Marcuzzi2016,Helmrich2019} and become balanced at a critical atom density $n_c\approx \Gamma/\kappa$. For $n_{\vec{x},t}>n_c$ (supercritical or active phase) individual excitations can grow into spatially extended clusters of excitations (avalanches) with a high degree of activity and particle loss. For $n_{\vec{x},t}<n_c$ (absorbing phase), on the other hand, excitation avalanches are rare or vanishingly small.

Figure~\ref{fig:soc_mechanism} illustrates the mechanisms leading to SOC. Starting from the supercritical regions of the cloud the density of excitations $\rho_{\vec{x},t}$ undergoes a period of rapid growth [labeled (i) in Fig.~\ref{fig:soc_mechanism}b], followed by a slow decrease in both $n_{\vec{x},t}$ and $\rho_{\vec{x},t}$ owing to a gradual loss of excited atoms [labeled (ii-a)]. In the limit of vanishingly small loss rate (perfect separation of timescales) the system will follow a characteristic trajectory  (dashed blue curve in Fig.~\ref{fig:soc_mechanism}b) that terminates at the critical point [orange cross at $n_{\vec{x},t}=n_c$ and $\rho_{\vec{x},t} = 0$]. On the other hand, if excitation avalanches persist on timescales comparable to the time for self-organization then the system dynamics may overshoot the critical point, terminating in the absorbing phase (dotted grey curve in Fig.~\ref{fig:soc_mechanism}b). This is associated with the appearance of a temporary dip in the atomic density distribution (grey curve in Fig.~\ref{fig:soc_mechanism}a). However the slow motion of particles in the trap refills this density dip, providing a refilling mechanism to escape the absorbing phase and approach the critical point [red curve in Fig.~\ref{fig:soc_mechanism}a, labeled (ii-b)]. This interplay of nonlinear excitation dynamics and atomic motion explains how the system self-organizes close to the critical point with a constant critical density across the cloud and sustains critical dynamics (e.g. avalanches) for long times compared to the initial self-organization period [labeled (iii)].

\begin{figure}
    \includegraphics[width=\linewidth]{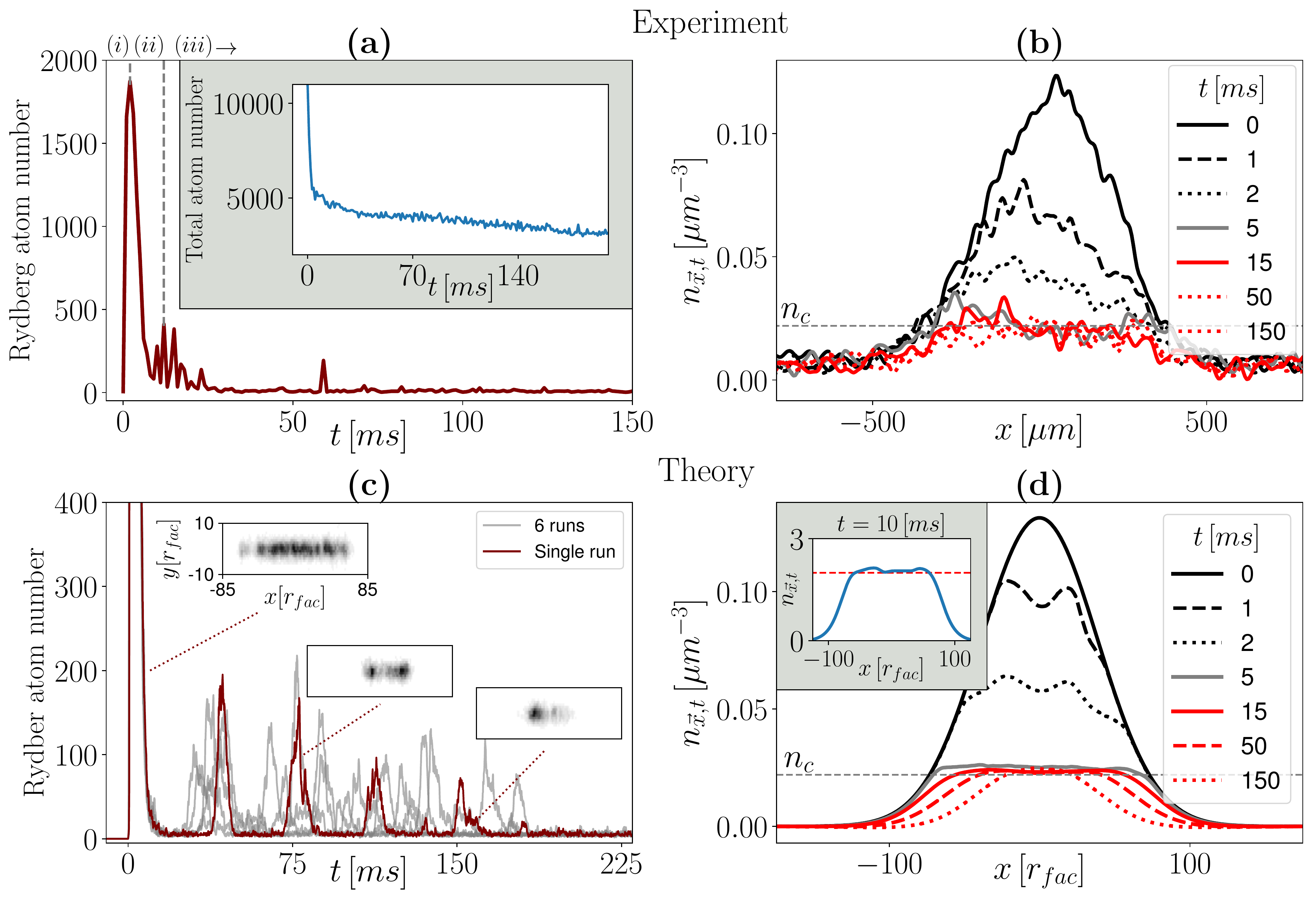}
    \caption{Theory-experiment comparison showing the approach to the SOC state (top: experiment, bottom: theory). (a) Instantaneous number of Rydberg excitations integrated over the cloud ($\propto\rho_t$). Each data point is obtained from a destructive measurement and corresponds to a distinct experimental realization. Inset: corresponding total atom number. (b) One-dimensional slices through the atomic density distribution $n_{\vec{x},t}$ at $y=z=0$. (c) Simulated dynamics of the full time-evolution (red line: single trajectory, grey lines: overlapped data of six different trajectories) showing temporally well separated, extensive excitation avalanches that persist long after the initial growth and self-organizing regimes (i) and (ii). (insets: snapshots of the peak excitation density per avalanche at $z=0$). (d) Slices through the simulated density profiles at $y=z=0$ showing the formation of a flat-top density profile pinned at $n_{\vec{x},t}=n_c$ analogous to the experimental observations in (b) (see also inset).}
    \label{fig:three_regimes}
\end{figure}

{\it Experimental approach.}--
Our experiments start with an ultracold gas of  $N=10^5$ potassium-39 atoms trapped in a cigar-shaped optical potential with trap frequencies of $\omega_{x}/{2\pi}=65\,\mathrm{Hz}$ and $\omega_{y,z}/{2\pi}=950\,\mathrm{Hz}$. The atomic cloud has a temperature of $T=40\,\mathrm{\mu K}$ and $e^{-1/2}$ radii of $\sigma_x=210\,\mathrm{\mu m}$, $\sigma_{y,z}=12\,\mathrm{\mu m}$ with a peak density of $n_0= 0.21\,\mathrm{\mu m}^{-3}$. 
At time $t=0$, we switch on an off-resonant ultraviolet (UV) laser coupling with Rabi frequency $\Omega\approx200$kHz and detuning $\Delta/ 2\pi = 30\,\mathrm{MHz}$ on the transition from the ground state $\ket{g}=\ket{4s_{1/2},F=1}$ to the Rydberg state $\ket{r}=\ket{66p_{3/2}}$. To strongly suppress single-particle excitations and to ensure that many-body effects dominate, we stay in the regime $\Gamma\ll \Omega \ll \Delta$~\cite{Lee2012,Weimer2012,Heidemann2007,Urvoy2015a,Weimer2008,Ates2007a,Amthor2010,Gart2013,Simonelli2016,Morsch2016a,Lesa2014,Morsch2016,Pupillo2010}.
Excitations decay with a calculated rate of $\Gamma/ 2\pi = 0.84 \,\mathrm{kHz}$, which either brings them back to the ground state $\ket{g}$ or into states $\ket{0}$ which are not coupled or are lost from the trap. This loss of particles  into inactive states $\ket{r}\rightarrow\ket{0}$ provides the crucial feedback mechanism for SOC~\cite{Helmrich2019}. 
After the laser exposure time $t$, we measure both the number of Rydberg excitations in the cloud as well as the spatial distribution of ground state atoms remaining in the trap. For the former we first ionize the Rydberg atoms and detect them on a micro-channel plate detector. For the latter we take an absorption image of the atom cloud, which integrates along the propagation of the light field~\cite{Helmrich2019}. 

Example absorption images after different exposure times $t$ are shown at the top of Fig.~\ref{fig:soc_mechanism}c, roughly coinciding with the ones sketched in Fig.~\ref{fig:soc_mechanism}a. The line profiles shown in Fig.~\ref{fig:soc_mechanism}c are reconstructed cross-sections of the three-dimensional density distribution through the center of the cloud. These are obtained by an inverse Abel transformation~\cite{hickstein2019PyAbel}, using radial symmetry along the elongated axis of the three-dimensional cigar-shaped cloud. Initially at $t=1\,\mathrm{ms}$, the gas has an approximate Gaussian shape as expected for an optically trapped thermal gas. At $t=5\,\mathrm{ms}$, a dip in the center where the density was initially the highest has developed. For even longer times $t\gtrapprox 15$\,ms, this dip has filled in and the cloud shows a flat-top coinciding with the critical density $n_c$. 

{\it Theoretical description.}-- The collective dynamics of the driven Rydberg ensemble is described by a nonequilibrium field theory for the local density of particles $n_{\vec{x},t}$ and the density of excitations $\rho_{\vec{x},t}$~\cite{Helmrich2019}. Besides the facilitated spreading of excitations and the dissipative decay, here we also account for the hydrodynamic motion of the atoms via two coupled stochastic evolution equations for $\rho_{\vec{x},t}, n_{\vec{x},t}$, including the internal and external degrees of freedom.

We label each atom with an index $j$, a set of operators $\sigma^{\alpha\beta}_j = \ket{\alpha}\bra{\beta}_j$ where $\alpha,\beta$ label the states $0,g,r$, and a position $\vec{x}_l$ (treated as classical variable). The equation of motion (EOM) for the internal degrees of freedom is given by the microscopic Liouvillian 
\begin{align}
    \partial_t \sigma^{\alpha\beta}_l&=i\left[ \biggl(\sum_{j\neq l} \frac{C_6\sigma_j^{rr}}{|\vec{x}_l-\vec{x}_j|^6}-\Delta\biggr)\sigma_l^{rr} + \Omega\frac{\sigma_l^{rg} + \sigma_l^{gr}}{2},\sigma^{\alpha\beta}_l\right]\nonumber\\
    &+\delta_{\alpha\beta}\left(\delta_{\alpha r}\gamma_{\text{de}}+\delta_{\alpha g} \gamma_{\downarrow g} + \delta_{\alpha 0}\gamma_{\downarrow 0}\right)\sigma^{\alpha\alpha}_l - \frac{\Gamma}{2}\{\sigma^{rr}_l,\sigma^{\alpha\beta}_l\},\nonumber
\end{align}
with the anti-commutator $\{\cdot,\cdot\}$, the commutator $[\cdot, \cdot]$ and the Kronecker symbol $\delta_{\alpha,\beta}$.
This includes coherent single-particle processes: laser driving with Rabi frequency $\Omega$ and detuning $\Delta$, and the van der Waals interaction between atoms $l$ and $j$ if both are in the Rydberg state. The dissipative single-particle processes are quantified by the dephasing rate $\gamma_{\text{de}}$,  the spontaneous decay rate $\gamma_{\downarrow g}$ for the process $\ket{r} \rightarrow \ket{g}$ ($\gamma_{\downarrow 0}$ for $\ket{r} \rightarrow \ket{0}$) and $\Gamma = \gamma_{\text{de}} + \gamma_{\downarrow g} + \gamma_{\downarrow 0}$.

In order to apply a coarse grained description, we define a unit cell as the sphere with radius $r_{\text{fac}}$ and volume $V_\text{fac}$. The densities per unit cell are~\cite{Helmrich2019}
${\rho_{\vec{x},t}=\sum_{j,\vec{x}}'\langle \sigma^{rr}_j \rangle / V_\text{fac}, n_{\vec{x},t}=\sum_{j,\vec{x}}'\langle \sigma^{rr}_j+\sigma^{gg}_j \rangle / V_\text{fac}}$ where  $\sum_{j,\vec{x}}'$ is restricted to $j\text{ with } |\vec{x}_j-\vec{x}|\le r_{\text{fac}}$. The EOM for the atomic density is evaluated by applying the chain rule
\begin{align}
\partial_t n_{\vec{x},t}=\sum_{j,\vec{x}}'\frac{\partial_t\langle \sigma^{rr}_j+\sigma^{gg}_j \rangle}{V_\text{fac}}-\nabla\sum_{j,\vec{x}}'\frac{\langle \sigma^{rr}_j+\sigma^{gg}_j \rangle}{V_\text{fac}}\partial_t\vec{x}_l, \label{EOM1}
\end{align}
where $\nabla=(\partial_{x},\partial_y,\partial_{z})$. It contains the EOM for the internal degrees of freedom and for the position of the atoms. The sum over the velocities in Eq.~\eqref{EOM1} is by definition the coarse grained current $\vec{j}$.

An equivalent computation for $\partial_t\vec{\rho}_{\vec{x},t}$ yields the Langevin equation~\cite{Marcuzzi2016,Buchhold2017}
\begin{eqnarray}
    \partial_t \rho_{\vec{x},t} &=& (D\nabla^2 - \Gamma)\rho_{\vec{x},t}+ (\tau+\kappa\rho_{\vec{x},t})\left(n_{\vec{x},t} - 2\rho_{\vec{x},t}\right) + \xi_{\vec{x},t}.\label{LangevinEq}
\end{eqnarray}
Here the evolution within each unit cell is decomposed in terms of facilitated (de-)excitation with rate $ \kappa\rho_{\vec{x},t}\approx \frac{\Omega^2 V_{\text{fac}}}{2\Delta}\rho_{\vec{x},t}$ and dissipative decay $\sim\Gamma$. Excitations spread diffusively between unit cells with $D r_{\text{fac}}\approx \kappa $. Rare, off-resonant single-particle excitations occur with rate $\tau r_{\text{fac}}^3=\frac{\kappa\Gamma}{\Delta}\approx10^{-4}\kappa$, acting as local seeds to prevent the system from getting stuck in an absorbing state. Local fluctuations in the excitation density are described by a multiplicative Markovian noise $\xi(\vec{x},t)$ with auto-correlation function $
    \langle \xi(\vec{x},t) \xi(\vec{y},t') \rangle = \delta(\vec{x}-\vec{y})\delta(t-t')\left(\Gamma\rho_{\vec{x},t} + \tau\right)$~\cite{Buchhold2017}.

The EOM of the density $n_{\vec{x},t}$ yields
\begin{eqnarray}
    \partial_t n_{\vec{x},t} &=& -\nabla\vec{j}_{\vec{x},t}-\gamma_{\downarrow 0}\rho_{\vec{x},t},\label{ContEq}
\end{eqnarray}
where the current $\vec{j}_{\vec{x},t}=-(D_T\nabla +\eta  \nabla V_{\vec{x}})n_{\vec{x},t}$ fits the common hydrodynamic form~\cite{supp_noto4}. It includes diffusion ($\propto D_T$) and an external force $-\nabla V_{\vec{x}}$ caused by the harmonic trapping potential $V_{\vec{x}} = \frac{M}{2}\sum_{l=x,y,z}(\omega_l \vec{x}_l)^2$, for which we use the frequencies $\omega_l$  and the atom mass $M$ from the experiment. The mobility $\eta$ is related to the diffusion constant $D_T = \eta k_B T$ via the Einstein relation. In the limit $\gamma_{\downarrow 0}\rightarrow0$, the steady state has zero current $\vec{j}_{\vec{x},t}=0$ and follows the Gaussian equilibrium distribution $n_{\vec{x},t} =n
^{(\text{eq})}_{\vec{x}}=n_0\exp(-\frac{V_{\vec{x}}}{k_BT})$. For the numerical simulation we use the trap frequencies $\omega_l$ and temperature $T$ measured in the experiment, and the initial spatial extension of the cloud is $\sigma_l=\sqrt{\frac{k_BT}{M\omega_l^2}}$, i.e.,  $\sigma_z=\sigma_y=2.5r_{\text{fac}}$ and $\sigma_x=44r_{\text{fac}}$. 
For $\gamma_{\downarrow 0}>0$ the current counteracts the loss and pulls density towards the center of the cloud, trying to rethermalize back to a Gaussian distribution with a typical rate $\eta M\omega_x^2$. In the EOM for $\rho_{\vec{x},t}$, particle motion is negligible compared to the facilitated spreading of excitations, i.e., $\sim D r_{\text{fac}}^{-2}\gg\eta M\omega_x^2$.

The Langevin equations are integrated numerically on a 3+1-dimensional lattice by means of an operator splitting scheme \cite{Dornic,Klocke,supp_noto4}, which remains well behaved in the limit $(\rho_{\vec{x},t},n_{\vec{x},t})\rightarrow(0,n_c)$ as well as at large system sizes and long times.
The parameters used in the simulations are chosen to match the experimentally observed facilitation and decay rates~\cite{Helmrich2019,noto3} as well as the real-space extension of the cloud. The mobility $\eta$ is hard to quantify from microscopic parameters alone and it was chosen such that the theoretical and experimental thermalization times match. This respects the separation of time scales between the fast spreading of excitations $\sim r_{\text{fac}}
^2/D=0.1$\,ms and the slow atomic motion in the trap $\sim 1/(\eta M\omega_x^2)=7$\,ms.

{\it Dynamics.}-- Figure~\ref{fig:three_regimes} shows comparable experiments and numerical simulations for an initially Gaussian atomic cloud with peak density $n>n_c$. Looking at both components $n_{\vec{x},t},\; \rho_{\vec{x},t}$ provides insights into the different dynamical regimes. This includes an initial growth regime (i), covering the first few milliseconds of evolution and resulting in a macroscopic Rydberg population. 
The early time growth dynamics are interesting in their own right~\cite{Wintermantel2020}, but are not overly important for the self-organizing behavior on longer timescales, apart from bringing the system into the initially supercritical phase. Subsequently the loss from the Rydberg state begins to decrease the total atom number. This leads to a self-organizing regime (ii), evidenced by large bursts of Rydberg excitations (large activity seen in Fig.~\ref{fig:three_regimes}a,c) and a sudden drop in the central density of the atomic cloud. The density approaches a flat-top distribution with a central density given by the critical value $n_{\vec{x},t}\approx n_c$ ($\approx 1/5$ of the initial peak density for our setup). This marks the onset of the self-organized critical regime (iii), characterised by an approximately flat central density $n_{\vec{x},t}=n_c$ (red curves in Fig.~\ref{fig:three_regimes}b,d) and sporadic avalanche-like excitation events. This is reached after approximately 15\,ms in the experiment and persists until at least 150\,ms. Simulations show that subsequent avalanches are well separated in space and time, which implies that the experimentally observed Rydberg excitation spikes correspond to individual avalanche events (Fig.~\ref{fig:three_regimes}c). In this regime each avalanche event transiently imprints a slight extra depression in the density profile such that $n_{\vec{x},t}< n_c$. However, particle transport from the flanks re-establishes $n_{\vec{x},t}\approx n_c$ between successive avalanches (Fig.~\ref{fig:three_regimes}) and pins $n_{\vec{x},t}=n_c$  (witnessed by Fig.~\ref{fig:three_regimes}d) thus sustaining a close to ideal critical SOC state over a large region of the system. 

In order to quantify the characteristic timescale associated to this mechanism, we investigate the effective refilling rate of the central region $\lambda$. A necessary condition for maintaining a SOC state is to satisfy a common separation of timescales $\gamma_{\downarrow 0} \gg \lambda \gg \tau$~\cite{Aschwanden2016,Klocke}. The refilling rate is determined by the gradient of the particle current $\lambda n_{\vec{x},t}\equiv-\nabla\vec{j}_{\vec{x},t}$ from the wings towards the center. In order to estimate $\lambda$, we apply a mean-field approach based on our observation that the current is dominated by particle flow along the elongated $x$-direction. Therefore, we assume a quasi-one-dimensional cloud with a flat-top of length $L_x$ and a constant density $n_{x,t} = \bar{n}_t \geq n_c$ inside the center. Outside, the density is sub-critical and follows the equilibrium profile $n_{x,t} = \bar n_t\left(n_x^{\rm (eq)} / n_{L_x/2}^{\rm (eq)}\right)$, which minimizes the current $\vec{j}_{x,t} = 0$ in the absence of excitations ($\rho_{x,t}=0$).
Averaging the current induced particle gain over the center yields
\begin{equation}
    \bar n_t\lambda=-\frac{1}{L_x}\int_{|x|\le L_x/2}\hspace{-4mm}dx\ \partial_x j_{x,t}=\eta M \omega_x^2\bar n_t.\label{eq:reloading}
\end{equation}
Using Eq.~\eqref{eq:reloading}, we estimate $\gamma_{\downarrow0}/\lambda\approx 50$ and $\lambda/\tau\approx 100$ from experimental parameters (and comparable for the  theory~\cite{noto3}). This indeed ensures both that individual avalanches experience a nearly constant central density over their lifetime ($\sim 1/\gamma_{\downarrow0}$). It also ensures that the refilling of the central density happens much faster than off-resonant excitations ($\sim1/\tau$), leading to well-separated avalanches, fulfilling the necessary conditions for SOC~\cite{Aschwanden2016}.
 
\begin{figure}
    \includegraphics[width=\linewidth]{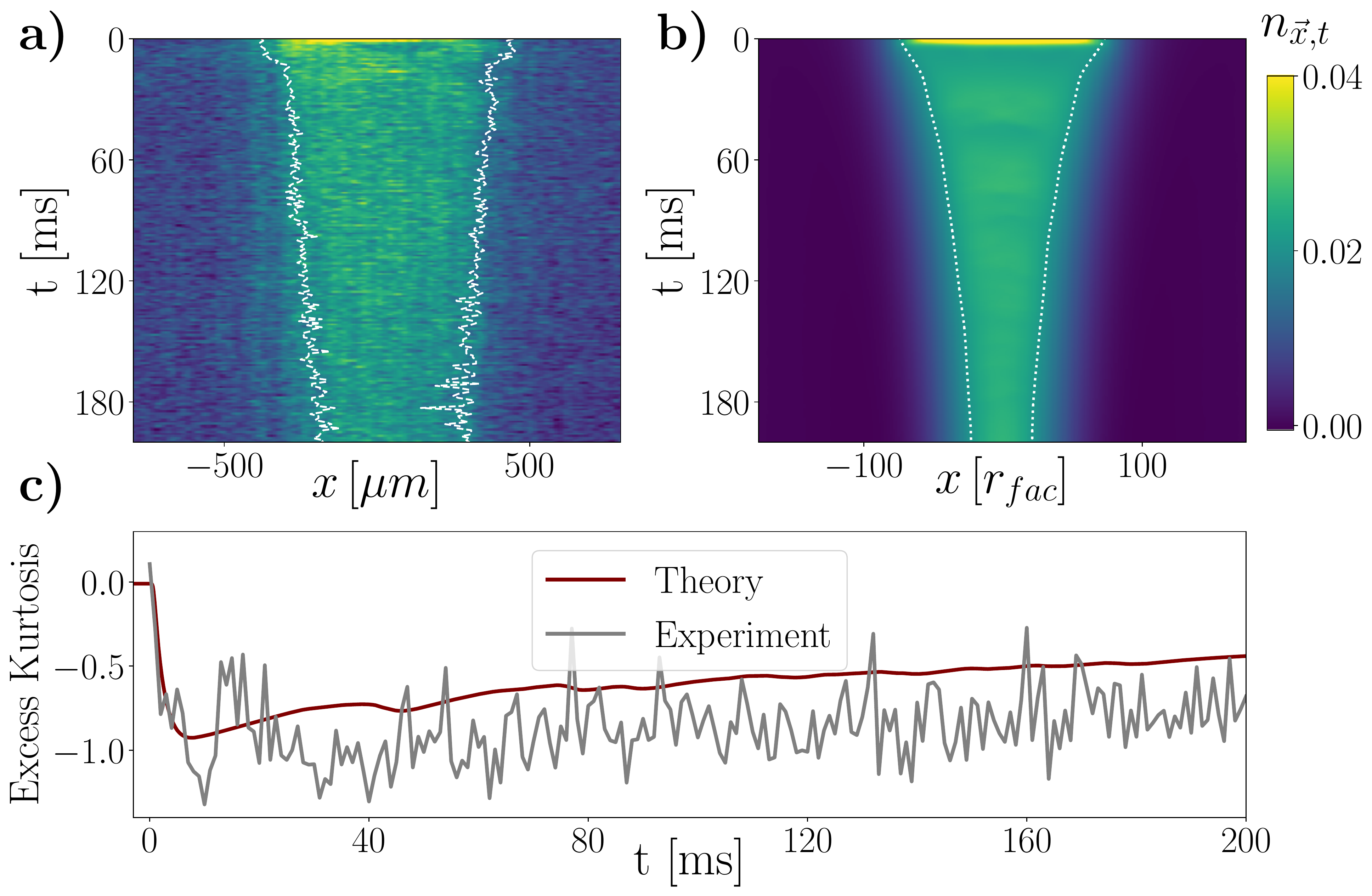}
    \caption{Melting of the flat-top from the time-evolved density profile $n_{\vec{x},t}$ (projected onto the $x$-axis). (a) Experimental measurements at different times and (b) the simulated evolution show a stable flat-top with a lifetime exceeding $200$\,ms and a boundary (white dashed curve), which slowly approaches the center. Both plots extend over the same $x$-axis distance. (c) The equilibration of the cloud profile is quantified by its time-dependent excess kurtosis EK$_t=\int dx (x/\sigma_x)^4 n_{\vec{x},t}-3$, where we integrate over a density slice with $y=z=0$. Here $\sigma_x$ is the width of the cloud in the $x$-direction. Starting from a Gaussian shape ($\text{EK}=0$), the kurtosis drops to $\text{EK}\approx-1$ after the initial avalanche. It recovers when the cloud evolves back towards an asymptotic thermal, Gaussian state.
    } \label{fig:profile_width}
\end{figure}

Finally, we analyze the very late time dynamics, in which the cloud reaches thermal equilibrium. The thermal state is approached when the particle reservoir represented by the flanks is continuously depleted, which in turn leads to the melting of the flat-top region (Fig.~\ref{fig:profile_width}a,b). The approach to thermal equilibrium can be seen by the evolution of the excess kurtosis EK$_t$, shown in Fig.~\ref{fig:profile_width}c. A non-zero kurtosis serves as a measure for the deviation of the cloud shape from a thermal Gaussian distribution, i.e., it measures relative flatness of the distribution. Its relaxation monitors the melting of the flat-top towards a robust, thermal equilibrium state without excitation outbursts (corresponding to EK$_t=0$). As seen in Fig.~\ref{fig:profile_width}c, the thermalization time scale exceeds $200$\,ms. Consequently we infer the lifetime of the SOC state to be at least $10$ times longer than the timescale associated with self-organization ($\approx20$\,ms). 

{\it Conclusion.}-- 
We have identified an important additional mechanism which explains how SOC can be sustained in a driven-dissipative ultracold atomic gas by nonequilibrium currents. We show that this generates a flat-top density distribution at the SOC critical density, quantitatively confirmed by the hydrodynamic Langevin theory. This demonstrates an important additional signature for SOC that could help identify SOC-like behavior in other systems, such as room-temperature atomic vapours and cold molecular plasmas~\cite{SOCAdams, wang2020}. Similar mechanisms may also be at play in very different systems including adaptive neural networks~\cite{Levina2007,Zierenberg}. The fact that the system naturally evolves to a stable, mostly homogeneous shape combined with the effectiveness of the hydrodynamic Langevin theory will enable more stringent tests of non-equilibrium universality in SOC systems. 
Alternatively, the interplay between internal and external degrees of freedom could lead to other rich dynamical regimes to test, such as oscillatory behaviour associated with SOqC~\cite{Bonachela2009,Bonachela2010,buenda2020feedback}. 

\begin{acknowledgments}
  \section*{Acknowledgements}
  K.K. acknowledges support from the National Science Foundation through grant DMR-1723367. T.M.W. acknowledges the French National Research Agency (ANR) through the Programme d'Investissement d'Avenir under contract ANR-17-EURE-0024. This project is part of and supported by DFG SPP 1929 GiRyd through projects DI1745/2-1 and WH141/3-3. S.W. is supported by the `Investissements d'Avenir' programme through the Excellence Initiative of the University of Strasbourg (IdEx) and the University of Strasbourg Institute for Advanced Study (USIAS). M.B. acknowledges support from the Alexander von Humboldt foundation.
\end{acknowledgments}
\bibliography{soc}

\begin{thebibliography}{61}%
\makeatletter
\providecommand \@ifxundefined [1]{%
 \@ifx{#1\undefined}
}%
\providecommand \@ifnum [1]{%
 \ifnum #1\expandafter \@firstoftwo
 \else \expandafter \@secondoftwo
 \fi
}%
\providecommand \@ifx [1]{%
 \ifx #1\expandafter \@firstoftwo
 \else \expandafter \@secondoftwo
 \fi
}%
\providecommand \natexlab [1]{#1}%
\providecommand \enquote  [1]{``#1''}%
\providecommand \bibnamefont  [1]{#1}%
\providecommand \bibfnamefont [1]{#1}%
\providecommand \citenamefont [1]{#1}%
\providecommand \href@noop [0]{\@secondoftwo}%
\providecommand \href [0]{\begingroup \@sanitize@url \@href}%
\providecommand \@href[1]{\@@startlink{#1}\@@href}%
\providecommand \@@href[1]{\endgroup#1\@@endlink}%
\providecommand \@sanitize@url [0]{\catcode `\\12\catcode `\$12\catcode
  `\&12\catcode `\#12\catcode `\^12\catcode `\_12\catcode `\%12\relax}%
\providecommand \@@startlink[1]{}%
\providecommand \@@endlink[0]{}%
\providecommand \url  [0]{\begingroup\@sanitize@url \@url }%
\providecommand \@url [1]{\endgroup\@href {#1}{\urlprefix }}%
\providecommand \urlprefix  [0]{URL }%
\providecommand \Eprint [0]{\href }%
\providecommand \doibase [0]{http://dx.doi.org/}%
\providecommand \selectlanguage [0]{\@gobble}%
\providecommand \bibinfo  [0]{\@secondoftwo}%
\providecommand \bibfield  [0]{\@secondoftwo}%
\providecommand \translation [1]{[#1]}%
\providecommand \BibitemOpen [0]{}%
\providecommand \bibitemStop [0]{}%
\providecommand \bibitemNoStop [0]{.\EOS\space}%
\providecommand \EOS [0]{\spacefactor3000\relax}%
\providecommand \BibitemShut  [1]{\csname bibitem#1\endcsname}%
\let\auto@bib@innerbib\@empty
\bibitem [{\citenamefont {Janssen}(1981)}]{Janssen1981}%
  \BibitemOpen
  \bibfield  {author} {\bibinfo {author} {\bibfnamefont {H.~K.}\ \bibnamefont
  {Janssen}},\ }\href {\doibase 10.1007/BF01319549} {\bibfield  {journal}
  {\bibinfo  {journal} {Zeitschrift f{\"u}r Physik B Condensed Matter}\
  }\textbf {\bibinfo {volume} {42}},\ \bibinfo {pages} {151} (\bibinfo {year}
  {1981})}\BibitemShut {NoStop}%
\bibitem [{\citenamefont {Grassberger}(1983)}]{Grassberger1983}%
  \BibitemOpen
  \bibfield  {author} {\bibinfo {author} {\bibfnamefont {P.}~\bibnamefont
  {Grassberger}},\ }\href@noop {} {\bibfield  {journal} {\bibinfo  {journal}
  {Mathematical Biosciences}\ }\textbf {\bibinfo {volume} {63}},\ \bibinfo
  {pages} {157} (\bibinfo {year} {1983})}\BibitemShut {NoStop}%
\bibitem [{\citenamefont {Babadi}\ \emph {et~al.}(2015)\citenamefont {Babadi},
  \citenamefont {Demler},\ and\ \citenamefont {Knap}}]{Babadi2015}%
  \BibitemOpen
  \bibfield  {author} {\bibinfo {author} {\bibfnamefont {M.}~\bibnamefont
  {Babadi}}, \bibinfo {author} {\bibfnamefont {E.}~\bibnamefont {Demler}}, \
  and\ \bibinfo {author} {\bibfnamefont {M.}~\bibnamefont {Knap}},\ }\href
  {\doibase 10.1103/PhysRevX.5.041005} {\bibfield  {journal} {\bibinfo
  {journal} {Phys. Rev. X}\ }\textbf {\bibinfo {volume} {5}},\ \bibinfo {pages}
  {041005} (\bibinfo {year} {2015})}\BibitemShut {NoStop}%
\bibitem [{\citenamefont {Tsatsos}\ \emph {et~al.}(2016)\citenamefont
  {Tsatsos}, \citenamefont {Tavares}, \citenamefont {Cidrim}, \citenamefont
  {Fritsch}, \citenamefont {Caracanhas}, \citenamefont {dos Santos},
  \citenamefont {Barenghi},\ and\ \citenamefont {Bagnato}}]{Tsatsos2016}%
  \BibitemOpen
  \bibfield  {author} {\bibinfo {author} {\bibfnamefont {M.~C.}\ \bibnamefont
  {Tsatsos}}, \bibinfo {author} {\bibfnamefont {P.~E.}\ \bibnamefont
  {Tavares}}, \bibinfo {author} {\bibfnamefont {A.}~\bibnamefont {Cidrim}},
  \bibinfo {author} {\bibfnamefont {A.~R.}\ \bibnamefont {Fritsch}}, \bibinfo
  {author} {\bibfnamefont {M.~A.}\ \bibnamefont {Caracanhas}}, \bibinfo
  {author} {\bibfnamefont {F.~E.~A.}\ \bibnamefont {dos Santos}}, \bibinfo
  {author} {\bibfnamefont {C.~F.}\ \bibnamefont {Barenghi}}, \ and\ \bibinfo
  {author} {\bibfnamefont {V.~S.}\ \bibnamefont {Bagnato}},\ }\href {\doibase
  https://doi.org/10.1016/j.physrep.2016.02.003} {\bibfield  {journal}
  {\bibinfo  {journal} {Physics Reports}\ }\textbf {\bibinfo {volume} {622}},\
  \bibinfo {pages} {1 } (\bibinfo {year} {2016})}\BibitemShut {NoStop}%
\bibitem [{\citenamefont {Kadau}\ \emph {et~al.}(2016)\citenamefont {Kadau},
  \citenamefont {Schmitt}, \citenamefont {Wenzel}, \citenamefont {Wink},
  \citenamefont {Maier}, \citenamefont {Ferrier-Barbut},\ and\ \citenamefont
  {Pfau}}]{Kadau2016}%
  \BibitemOpen
  \bibfield  {author} {\bibinfo {author} {\bibfnamefont {H.}~\bibnamefont
  {Kadau}}, \bibinfo {author} {\bibfnamefont {M.}~\bibnamefont {Schmitt}},
  \bibinfo {author} {\bibfnamefont {M.}~\bibnamefont {Wenzel}}, \bibinfo
  {author} {\bibfnamefont {C.}~\bibnamefont {Wink}}, \bibinfo {author}
  {\bibfnamefont {T.}~\bibnamefont {Maier}}, \bibinfo {author} {\bibfnamefont
  {I.}~\bibnamefont {Ferrier-Barbut}}, \ and\ \bibinfo {author} {\bibfnamefont
  {T.}~\bibnamefont {Pfau}},\ }\href {\doibase 10.1038/nature16485} {\bibfield
  {journal} {\bibinfo  {journal} {Nature}\ }\textbf {\bibinfo {volume} {530}},\
  \bibinfo {pages} {194} (\bibinfo {year} {2016})}\BibitemShut {NoStop}%
\bibitem [{\citenamefont {Gillman}\ \emph {et~al.}(2020)\citenamefont
  {Gillman}, \citenamefont {Carollo},\ and\ \citenamefont
  {Lesanovsky}}]{gillman2020}%
  \BibitemOpen
  \bibfield  {author} {\bibinfo {author} {\bibfnamefont {E.}~\bibnamefont
  {Gillman}}, \bibinfo {author} {\bibfnamefont {F.}~\bibnamefont {Carollo}}, \
  and\ \bibinfo {author} {\bibfnamefont {I.}~\bibnamefont {Lesanovsky}},\
  }\href {\doibase 10.1103/PhysRevLett.125.100403} {\bibfield  {journal}
  {\bibinfo  {journal} {Phys. Rev. Lett.}\ }\textbf {\bibinfo {volume} {125}},\
  \bibinfo {pages} {100403} (\bibinfo {year} {2020})}\BibitemShut {NoStop}%
\bibitem [{\citenamefont {Mu\~noz}\ \emph {et~al.}(1999)\citenamefont
  {Mu\~noz}, \citenamefont {Dickman}, \citenamefont {Vespignani},\ and\
  \citenamefont {Zapperi}}]{Munoz1999}%
  \BibitemOpen
  \bibfield  {author} {\bibinfo {author} {\bibfnamefont {M.~A.}\ \bibnamefont
  {Mu\~noz}}, \bibinfo {author} {\bibfnamefont {R.}~\bibnamefont {Dickman}},
  \bibinfo {author} {\bibfnamefont {A.}~\bibnamefont {Vespignani}}, \ and\
  \bibinfo {author} {\bibfnamefont {S.}~\bibnamefont {Zapperi}},\ }\href
  {\doibase 10.1103/PhysRevE.59.6175} {\bibfield  {journal} {\bibinfo
  {journal} {Phys. Rev. E}\ }\textbf {\bibinfo {volume} {59}},\ \bibinfo
  {pages} {6175} (\bibinfo {year} {1999})}\BibitemShut {NoStop}%
\bibitem [{\citenamefont {Takeuchi}\ \emph {et~al.}(2007)\citenamefont
  {Takeuchi}, \citenamefont {Kuroda}, \citenamefont {Chat{\'{e}}},\ and\
  \citenamefont {Sano}}]{Takeuchi2007}%
  \BibitemOpen
  \bibfield  {author} {\bibinfo {author} {\bibfnamefont {K.~A.}\ \bibnamefont
  {Takeuchi}}, \bibinfo {author} {\bibfnamefont {M.}~\bibnamefont {Kuroda}},
  \bibinfo {author} {\bibfnamefont {H.}~\bibnamefont {Chat{\'{e}}}}, \ and\
  \bibinfo {author} {\bibfnamefont {M.}~\bibnamefont {Sano}},\ }\href {\doibase
  10.1103/PhysRevLett.99.234503} {\bibfield  {journal} {\bibinfo  {journal}
  {Phys. Rev. Lett.}\ }\textbf {\bibinfo {volume} {99}},\ \bibinfo {pages}
  {234503} (\bibinfo {year} {2007})}\BibitemShut {NoStop}%
\bibitem [{\citenamefont {Lemoult}\ \emph {et~al.}(2016)\citenamefont
  {Lemoult}, \citenamefont {Shi}, \citenamefont {Avila}, \citenamefont
  {Jalikop}, \citenamefont {Avila},\ and\ \citenamefont {Hof}}]{Lemoult2016}%
  \BibitemOpen
  \bibfield  {author} {\bibinfo {author} {\bibfnamefont {G.}~\bibnamefont
  {Lemoult}}, \bibinfo {author} {\bibfnamefont {L.}~\bibnamefont {Shi}},
  \bibinfo {author} {\bibfnamefont {K.}~\bibnamefont {Avila}}, \bibinfo
  {author} {\bibfnamefont {S.~V.}\ \bibnamefont {Jalikop}}, \bibinfo {author}
  {\bibfnamefont {M.}~\bibnamefont {Avila}}, \ and\ \bibinfo {author}
  {\bibfnamefont {B.}~\bibnamefont {Hof}},\ }\href {\doibase 10.1038/nphys3675}
  {\bibfield  {journal} {\bibinfo  {journal} {Nature Physics}\ }\textbf
  {\bibinfo {volume} {12}},\ \bibinfo {pages} {254} (\bibinfo {year}
  {2016})}\BibitemShut {NoStop}%
\bibitem [{\citenamefont {Sano}\ and\ \citenamefont {Tamai}(2016)}]{Sano2016}%
  \BibitemOpen
  \bibfield  {author} {\bibinfo {author} {\bibfnamefont {M.}~\bibnamefont
  {Sano}}\ and\ \bibinfo {author} {\bibfnamefont {K.}~\bibnamefont {Tamai}},\
  }\href {http://dx.doi.org/10.1038/nphys3659 http://10.0.4.14/nphys3659
  https://www.nature.com/articles/nphys3659{\#}supplementary-information}
  {\bibfield  {journal} {\bibinfo  {journal} {Nature Physics}\ }\textbf
  {\bibinfo {volume} {12}},\ \bibinfo {pages} {249} (\bibinfo {year}
  {2016})}\BibitemShut {NoStop}%
\bibitem [{\citenamefont {Nowak}\ \emph {et~al.}(2011)\citenamefont {Nowak},
  \citenamefont {Sexty},\ and\ \citenamefont {Gasenzer}}]{Nowak}%
  \BibitemOpen
  \bibfield  {author} {\bibinfo {author} {\bibfnamefont {B.}~\bibnamefont
  {Nowak}}, \bibinfo {author} {\bibfnamefont {D.}~\bibnamefont {Sexty}}, \ and\
  \bibinfo {author} {\bibfnamefont {T.}~\bibnamefont {Gasenzer}},\ }\href
  {\doibase 10.1103/PhysRevB.84.020506} {\bibfield  {journal} {\bibinfo
  {journal} {Phys. Rev. B}\ }\textbf {\bibinfo {volume} {84}},\ \bibinfo
  {pages} {020506} (\bibinfo {year} {2011})}\BibitemShut {NoStop}%
\bibitem [{\citenamefont {Bak}\ \emph {et~al.}(1987)\citenamefont {Bak},
  \citenamefont {Tang},\ and\ \citenamefont {Wiesenfeld}}]{Bak1987}%
  \BibitemOpen
  \bibfield  {author} {\bibinfo {author} {\bibfnamefont {P.}~\bibnamefont
  {Bak}}, \bibinfo {author} {\bibfnamefont {C.}~\bibnamefont {Tang}}, \ and\
  \bibinfo {author} {\bibfnamefont {K.}~\bibnamefont {Wiesenfeld}},\ }\href
  {\doibase 10.1103/PhysRevLett.59.381} {\bibfield  {journal} {\bibinfo
  {journal} {Phys. Rev. Lett.}\ }\textbf {\bibinfo {volume} {59}},\ \bibinfo
  {pages} {381} (\bibinfo {year} {1987})}\BibitemShut {NoStop}%
\bibitem [{\citenamefont {Buchhold}\ and\ \citenamefont
  {Diehl}(2015)}]{buchhold2015}%
  \BibitemOpen
  \bibfield  {author} {\bibinfo {author} {\bibfnamefont {M.}~\bibnamefont
  {Buchhold}}\ and\ \bibinfo {author} {\bibfnamefont {S.}~\bibnamefont
  {Diehl}},\ }\href {\doibase 10.1103/PhysRevA.92.013603} {\bibfield  {journal}
  {\bibinfo  {journal} {Phys. Rev. A}\ }\textbf {\bibinfo {volume} {92}},\
  \bibinfo {pages} {013603} (\bibinfo {year} {2015})}\BibitemShut {NoStop}%
\bibitem [{\citenamefont {Berges}\ \emph {et~al.}(2008)\citenamefont {Berges},
  \citenamefont {Rothkopf},\ and\ \citenamefont {Schmidt}}]{Berges2008}%
  \BibitemOpen
  \bibfield  {author} {\bibinfo {author} {\bibfnamefont {J.}~\bibnamefont
  {Berges}}, \bibinfo {author} {\bibfnamefont {A.}~\bibnamefont {Rothkopf}}, \
  and\ \bibinfo {author} {\bibfnamefont {J.}~\bibnamefont {Schmidt}},\ }\href
  {\doibase 10.1103/PhysRevLett.101.041603} {\bibfield  {journal} {\bibinfo
  {journal} {Phys. Rev. Lett.}\ }\textbf {\bibinfo {volume} {101}},\ \bibinfo
  {pages} {041603} (\bibinfo {year} {2008})}\BibitemShut {NoStop}%
\bibitem [{\citenamefont {{Kinouchi}}\ and\ \citenamefont
  {{Copelli}}(2006)}]{Kinouchi2006}%
  \BibitemOpen
  \bibfield  {author} {\bibinfo {author} {\bibfnamefont {O.}~\bibnamefont
  {{Kinouchi}}}\ and\ \bibinfo {author} {\bibfnamefont {M.}~\bibnamefont
  {{Copelli}}},\ }\href {\doibase 10.1038/nphys289} {\bibfield  {journal}
  {\bibinfo  {journal} {Nature Physics}\ }\textbf {\bibinfo {volume} {2}},\
  \bibinfo {pages} {348} (\bibinfo {year} {2006})}\BibitemShut {NoStop}%
\bibitem [{\citenamefont {Bertschinger}\ and\ \citenamefont
  {Natschl\"ager}(2004)}]{Bertsch2004}%
  \BibitemOpen
  \bibfield  {author} {\bibinfo {author} {\bibfnamefont {N.}~\bibnamefont
  {Bertschinger}}\ and\ \bibinfo {author} {\bibfnamefont {T.}~\bibnamefont
  {Natschl\"ager}},\ }\href {\doibase 10.1162/089976604323057443} {\bibfield
  {journal} {\bibinfo  {journal} {Neural Computation}\ }\textbf {\bibinfo
  {volume} {16}},\ \bibinfo {pages} {1413} (\bibinfo {year}
  {2004})}\BibitemShut {NoStop}%
\bibitem [{\citenamefont {Bornholdt}\ and\ \citenamefont
  {Rohlf}(2000)}]{Bornholdt2000}%
  \BibitemOpen
  \bibfield  {author} {\bibinfo {author} {\bibfnamefont {S.}~\bibnamefont
  {Bornholdt}}\ and\ \bibinfo {author} {\bibfnamefont {T.}~\bibnamefont
  {Rohlf}},\ }\href {\doibase 10.1103/PhysRevLett.84.6114} {\bibfield
  {journal} {\bibinfo  {journal} {Phys. Rev. Lett.}\ }\textbf {\bibinfo
  {volume} {84}},\ \bibinfo {pages} {6114} (\bibinfo {year}
  {2000})}\BibitemShut {NoStop}%
\bibitem [{\citenamefont {Nicklas}\ \emph {et~al.}(2015)\citenamefont
  {Nicklas}, \citenamefont {Karl}, \citenamefont {H{\"{o}}fer}, \citenamefont
  {Johnson}, \citenamefont {Muessel}, \citenamefont {Strobel}, \citenamefont
  {{Tomkovi{\v{c}}fi}}, \citenamefont {Gasenzer},\ and\ \citenamefont
  {Oberthaler}}]{Nicklas2015}%
  \BibitemOpen
  \bibfield  {author} {\bibinfo {author} {\bibfnamefont {E.}~\bibnamefont
  {Nicklas}}, \bibinfo {author} {\bibfnamefont {M.}~\bibnamefont {Karl}},
  \bibinfo {author} {\bibfnamefont {M.}~\bibnamefont {H{\"{o}}fer}}, \bibinfo
  {author} {\bibfnamefont {A.}~\bibnamefont {Johnson}}, \bibinfo {author}
  {\bibfnamefont {W.}~\bibnamefont {Muessel}}, \bibinfo {author} {\bibfnamefont
  {H.}~\bibnamefont {Strobel}}, \bibinfo {author} {\bibfnamefont
  {J.}~\bibnamefont {{Tomkovi{\v{c}}fi}}}, \bibinfo {author} {\bibfnamefont
  {T.}~\bibnamefont {Gasenzer}}, \ and\ \bibinfo {author} {\bibfnamefont
  {M.~K.}\ \bibnamefont {Oberthaler}},\ }\href {\doibase
  10.1103/PhysRevLett.115.245301} {\bibfield  {journal} {\bibinfo  {journal}
  {Phys. Rev. Lett.}\ }\textbf {\bibinfo {volume} {115}},\ \bibinfo {pages}
  {245301} (\bibinfo {year} {2015})}\BibitemShut {NoStop}%
\bibitem [{\citenamefont {Bak}\ \emph {et~al.}(1988)\citenamefont {Bak},
  \citenamefont {Tang},\ and\ \citenamefont {Wiesenfeld}}]{Bak1988}%
  \BibitemOpen
  \bibfield  {author} {\bibinfo {author} {\bibfnamefont {P.}~\bibnamefont
  {Bak}}, \bibinfo {author} {\bibfnamefont {C.}~\bibnamefont {Tang}}, \ and\
  \bibinfo {author} {\bibfnamefont {K.}~\bibnamefont {Wiesenfeld}},\ }\href
  {\doibase 10.1103/PhysRevA.38.364} {\bibfield  {journal} {\bibinfo  {journal}
  {Phys. Rev. A}\ }\textbf {\bibinfo {volume} {38}},\ \bibinfo {pages} {364}
  (\bibinfo {year} {1988})}\BibitemShut {NoStop}%
\bibitem [{\citenamefont {Altshuler}\ and\ \citenamefont
  {Johansen}(2004)}]{Altshuler2004}%
  \BibitemOpen
  \bibfield  {author} {\bibinfo {author} {\bibfnamefont {E.}~\bibnamefont
  {Altshuler}}\ and\ \bibinfo {author} {\bibfnamefont {T.~H.}\ \bibnamefont
  {Johansen}},\ }\href {\doibase 10.1103/RevModPhys.76.471} {\bibfield
  {journal} {\bibinfo  {journal} {Rev. Mod. Phys.}\ }\textbf {\bibinfo {volume}
  {76}},\ \bibinfo {pages} {471} (\bibinfo {year} {2004})}\BibitemShut
  {NoStop}%
\bibitem [{\citenamefont {Field}\ \emph {et~al.}(1995)\citenamefont {Field},
  \citenamefont {Witt}, \citenamefont {Nori},\ and\ \citenamefont
  {Ling}}]{Field1995}%
  \BibitemOpen
  \bibfield  {author} {\bibinfo {author} {\bibfnamefont {S.}~\bibnamefont
  {Field}}, \bibinfo {author} {\bibfnamefont {J.}~\bibnamefont {Witt}},
  \bibinfo {author} {\bibfnamefont {F.}~\bibnamefont {Nori}}, \ and\ \bibinfo
  {author} {\bibfnamefont {X.}~\bibnamefont {Ling}},\ }\href {\doibase
  10.1103/PhysRevLett.74.1206} {\bibfield  {journal} {\bibinfo  {journal}
  {Phys. Rev. Lett.}\ }\textbf {\bibinfo {volume} {74}},\ \bibinfo {pages}
  {1206} (\bibinfo {year} {1995})}\BibitemShut {NoStop}%
\bibitem [{\citenamefont {Jian}\ \emph {et~al.}(2019)\citenamefont {Jian},
  \citenamefont {Yin},\ and\ \citenamefont {Swingle}}]{Swingle}%
  \BibitemOpen
  \bibfield  {author} {\bibinfo {author} {\bibfnamefont {S.-K.}\ \bibnamefont
  {Jian}}, \bibinfo {author} {\bibfnamefont {S.}~\bibnamefont {Yin}}, \ and\
  \bibinfo {author} {\bibfnamefont {B.}~\bibnamefont {Swingle}},\ }\href
  {\doibase 10.1103/PhysRevLett.123.170606} {\bibfield  {journal} {\bibinfo
  {journal} {Phys. Rev. Lett.}\ }\textbf {\bibinfo {volume} {123}},\ \bibinfo
  {pages} {170606} (\bibinfo {year} {2019})}\BibitemShut {NoStop}%
\bibitem [{\citenamefont {Aschwanden}\ \emph {et~al.}(2016)\citenamefont
  {Aschwanden}, \citenamefont {Crosby}, \citenamefont {Dimitropoulou},
  \citenamefont {Georgoulis}, \citenamefont {Hergarten}, \citenamefont
  {McAteer}, \citenamefont {Milovanov}, \citenamefont {Mineshige},
  \citenamefont {Morales}, \citenamefont {Nishizuka}, \citenamefont
  {Pruessner}, \citenamefont {Sanchez}, \citenamefont {Sharma}, \citenamefont
  {Strugarek},\ and\ \citenamefont {Uritsky}}]{Aschwanden2016}%
  \BibitemOpen
  \bibfield  {author} {\bibinfo {author} {\bibfnamefont {M.~J.}\ \bibnamefont
  {Aschwanden}}, \bibinfo {author} {\bibfnamefont {N.~B.}\ \bibnamefont
  {Crosby}}, \bibinfo {author} {\bibfnamefont {M.}~\bibnamefont
  {Dimitropoulou}}, \bibinfo {author} {\bibfnamefont {M.~K.}\ \bibnamefont
  {Georgoulis}}, \bibinfo {author} {\bibfnamefont {S.}~\bibnamefont
  {Hergarten}}, \bibinfo {author} {\bibfnamefont {J.}~\bibnamefont {McAteer}},
  \bibinfo {author} {\bibfnamefont {A.~V.}\ \bibnamefont {Milovanov}}, \bibinfo
  {author} {\bibfnamefont {S.}~\bibnamefont {Mineshige}}, \bibinfo {author}
  {\bibfnamefont {L.}~\bibnamefont {Morales}}, \bibinfo {author} {\bibfnamefont
  {N.}~\bibnamefont {Nishizuka}}, \bibinfo {author} {\bibfnamefont
  {G.}~\bibnamefont {Pruessner}}, \bibinfo {author} {\bibfnamefont
  {R.}~\bibnamefont {Sanchez}}, \bibinfo {author} {\bibfnamefont {A.~S.}\
  \bibnamefont {Sharma}}, \bibinfo {author} {\bibfnamefont {A.}~\bibnamefont
  {Strugarek}}, \ and\ \bibinfo {author} {\bibfnamefont {V.}~\bibnamefont
  {Uritsky}},\ }\href {\doibase 10.1007/s11214-014-0054-6} {\bibfield
  {journal} {\bibinfo  {journal} {Space Science Reviews}\ }\textbf {\bibinfo
  {volume} {198}},\ \bibinfo {pages} {47} (\bibinfo {year} {2016})}\BibitemShut
  {NoStop}%
\bibitem [{\citenamefont {{Turcotte}}(1999)}]{Turcotte1999}%
  \BibitemOpen
  \bibfield  {author} {\bibinfo {author} {\bibfnamefont {D.~L.}\ \bibnamefont
  {{Turcotte}}},\ }\href {\doibase 10.1088/0034-4885/62/10/201} {\bibfield
  {journal} {\bibinfo  {journal} {Reports on Progress in Physics}\ }\textbf
  {\bibinfo {volume} {62}},\ \bibinfo {pages} {1377} (\bibinfo {year}
  {1999})}\BibitemShut {NoStop}%
\bibitem [{\citenamefont {Pruessner}(2012)}]{pruessner2012self}%
  \BibitemOpen
  \bibfield  {author} {\bibinfo {author} {\bibfnamefont {G.}~\bibnamefont
  {Pruessner}},\ }\href@noop {} {\emph {\bibinfo {title} {Self-organised
  criticality: theory, models and characterisation}}}\ (\bibinfo  {publisher}
  {Cambridge University Press},\ \bibinfo {year} {2012})\BibitemShut {NoStop}%
\bibitem [{\citenamefont {Bonachela}\ and\ \citenamefont
  {Mu{\~{n}}oz}(2009)}]{Bonachela2009}%
  \BibitemOpen
  \bibfield  {author} {\bibinfo {author} {\bibfnamefont {J.~A.}\ \bibnamefont
  {Bonachela}}\ and\ \bibinfo {author} {\bibfnamefont {M.~A.}\ \bibnamefont
  {Mu{\~{n}}oz}},\ }\href {\doibase 10.1088/1742-5468/2009/09/P09009}
  {\bibfield  {journal} {\bibinfo  {journal} {Journal of Statistical Mechanics:
  Theory and Experiment}\ }\textbf {\bibinfo {volume} {2009}},\ \bibinfo
  {pages} {P09009} (\bibinfo {year} {2009})}\BibitemShut {NoStop}%
\bibitem [{\citenamefont {Bonachela}\ \emph {et~al.}(2010)\citenamefont
  {Bonachela}, \citenamefont {de~Franciscis}, \citenamefont {Torres},\ and\
  \citenamefont {Mu{\~{n}}oz}}]{Bonachela2010}%
  \BibitemOpen
  \bibfield  {author} {\bibinfo {author} {\bibfnamefont {J.~A.}\ \bibnamefont
  {Bonachela}}, \bibinfo {author} {\bibfnamefont {S.}~\bibnamefont
  {de~Franciscis}}, \bibinfo {author} {\bibfnamefont {J.~J.}\ \bibnamefont
  {Torres}}, \ and\ \bibinfo {author} {\bibfnamefont {M.~A.}\ \bibnamefont
  {Mu{\~{n}}oz}},\ }\href {http://stacks.iop.org/1742-5468/2010/i=02/a=P02015}
  {\bibfield  {journal} {\bibinfo  {journal} {Journal of Statistical Mechanics:
  Theory and Experiment}\ }\textbf {\bibinfo {volume} {2010}},\ \bibinfo
  {pages} {P02015} (\bibinfo {year} {2010})}\BibitemShut {NoStop}%
\bibitem [{\citenamefont {Helmrich}\ \emph {et~al.}(2020)\citenamefont
  {Helmrich}, \citenamefont {Arias}, \citenamefont {Lochead}, \citenamefont
  {Wintermantel}, \citenamefont {Buchhold}, \citenamefont {Diehl},\ and\
  \citenamefont {Whitlock}}]{Helmrich2019}%
  \BibitemOpen
  \bibfield  {author} {\bibinfo {author} {\bibfnamefont {S.}~\bibnamefont
  {Helmrich}}, \bibinfo {author} {\bibfnamefont {A.}~\bibnamefont {Arias}},
  \bibinfo {author} {\bibfnamefont {G.}~\bibnamefont {Lochead}}, \bibinfo
  {author} {\bibfnamefont {T.~M.}\ \bibnamefont {Wintermantel}}, \bibinfo
  {author} {\bibfnamefont {M.}~\bibnamefont {Buchhold}}, \bibinfo {author}
  {\bibfnamefont {S.}~\bibnamefont {Diehl}}, \ and\ \bibinfo {author}
  {\bibfnamefont {S.}~\bibnamefont {Whitlock}},\ }\href {\doibase
  10.1038/s41586-019-1908-6} {\bibfield  {journal} {\bibinfo  {journal}
  {Nature}\ }\textbf {\bibinfo {volume} {577}},\ \bibinfo {pages} {481}
  (\bibinfo {year} {2020})}\BibitemShut {NoStop}%
\bibitem [{\citenamefont {Ding}\ \emph {et~al.}(2020)\citenamefont {Ding},
  \citenamefont {Busche}, \citenamefont {Shi}, \citenamefont {Guo},\ and\
  \citenamefont {Adams}}]{SOCAdams}%
  \BibitemOpen
  \bibfield  {author} {\bibinfo {author} {\bibfnamefont {D.-S.}\ \bibnamefont
  {Ding}}, \bibinfo {author} {\bibfnamefont {H.}~\bibnamefont {Busche}},
  \bibinfo {author} {\bibfnamefont {B.-S.}\ \bibnamefont {Shi}}, \bibinfo
  {author} {\bibfnamefont {G.-C.}\ \bibnamefont {Guo}}, \ and\ \bibinfo
  {author} {\bibfnamefont {C.~S.}\ \bibnamefont {Adams}},\ }\href {\doibase
  10.1103/PhysRevX.10.021023} {\bibfield  {journal} {\bibinfo  {journal} {Phys.
  Rev. X}\ }\textbf {\bibinfo {volume} {10}},\ \bibinfo {pages} {021023}
  (\bibinfo {year} {2020})}\BibitemShut {NoStop}%
\bibitem [{\citenamefont {{Levina}}\ \emph {et~al.}(2007)\citenamefont
  {{Levina}}, \citenamefont {{Herrmann}},\ and\ \citenamefont
  {{Geisel}}}]{Levina2007}%
  \BibitemOpen
  \bibfield  {author} {\bibinfo {author} {\bibfnamefont {A.}~\bibnamefont
  {{Levina}}}, \bibinfo {author} {\bibfnamefont {J.~M.}\ \bibnamefont
  {{Herrmann}}}, \ and\ \bibinfo {author} {\bibfnamefont {T.}~\bibnamefont
  {{Geisel}}},\ }\href {\doibase 10.1038/nphys758} {\bibfield  {journal}
  {\bibinfo  {journal} {Nature Physics}\ }\textbf {\bibinfo {volume} {3}},\
  \bibinfo {pages} {857} (\bibinfo {year} {2007})}\BibitemShut {NoStop}%
\bibitem [{\citenamefont {Zierenberg}\ \emph {et~al.}(2018)\citenamefont
  {Zierenberg}, \citenamefont {Wilting},\ and\ \citenamefont
  {Priesemann}}]{Zierenberg}%
  \BibitemOpen
  \bibfield  {author} {\bibinfo {author} {\bibfnamefont {J.}~\bibnamefont
  {Zierenberg}}, \bibinfo {author} {\bibfnamefont {J.}~\bibnamefont {Wilting}},
  \ and\ \bibinfo {author} {\bibfnamefont {V.}~\bibnamefont {Priesemann}},\
  }\href {\doibase 10.1103/PhysRevX.8.031018} {\bibfield  {journal} {\bibinfo
  {journal} {Phys. Rev. X}\ }\textbf {\bibinfo {volume} {8}},\ \bibinfo {pages}
  {031018} (\bibinfo {year} {2018})}\BibitemShut {NoStop}%
\bibitem [{\citenamefont {Schempp}\ \emph {et~al.}(2014)\citenamefont
  {Schempp}, \citenamefont {G\"unter}, \citenamefont {Robert-de Saint-Vincent},
  \citenamefont {Hofmann}, \citenamefont {Breyel}, \citenamefont {Komnik},
  \citenamefont {Sch\"onleber}, \citenamefont {G\"arttner}, \citenamefont
  {Evers}, \citenamefont {Whitlock},\ and\ \citenamefont
  {Weidem\"uller}}]{Schempp2014}%
  \BibitemOpen
  \bibfield  {author} {\bibinfo {author} {\bibfnamefont {H.}~\bibnamefont
  {Schempp}}, \bibinfo {author} {\bibfnamefont {G.}~\bibnamefont {G\"unter}},
  \bibinfo {author} {\bibfnamefont {M.}~\bibnamefont {Robert-de
  Saint-Vincent}}, \bibinfo {author} {\bibfnamefont {C.~S.}\ \bibnamefont
  {Hofmann}}, \bibinfo {author} {\bibfnamefont {D.}~\bibnamefont {Breyel}},
  \bibinfo {author} {\bibfnamefont {A.}~\bibnamefont {Komnik}}, \bibinfo
  {author} {\bibfnamefont {D.~W.}\ \bibnamefont {Sch\"onleber}}, \bibinfo
  {author} {\bibfnamefont {M.}~\bibnamefont {G\"arttner}}, \bibinfo {author}
  {\bibfnamefont {J.}~\bibnamefont {Evers}}, \bibinfo {author} {\bibfnamefont
  {S.}~\bibnamefont {Whitlock}}, \ and\ \bibinfo {author} {\bibfnamefont
  {M.}~\bibnamefont {Weidem\"uller}},\ }\href {\doibase
  10.1103/PhysRevLett.112.013002} {\bibfield  {journal} {\bibinfo  {journal}
  {Phys. Rev. Lett.}\ }\textbf {\bibinfo {volume} {112}},\ \bibinfo {pages}
  {13002} (\bibinfo {year} {2014})}\BibitemShut {NoStop}%
\bibitem [{\citenamefont {Malossi}\ \emph {et~al.}(2014)\citenamefont
  {Malossi}, \citenamefont {Valado}, \citenamefont {Scotto}, \citenamefont
  {Huillery}, \citenamefont {Pillet}, \citenamefont {Ciampini}, \citenamefont
  {Arimondo},\ and\ \citenamefont {Morsch}}]{Malossi2014}%
  \BibitemOpen
  \bibfield  {author} {\bibinfo {author} {\bibfnamefont {N.}~\bibnamefont
  {Malossi}}, \bibinfo {author} {\bibfnamefont {M.~M.}\ \bibnamefont {Valado}},
  \bibinfo {author} {\bibfnamefont {S.}~\bibnamefont {Scotto}}, \bibinfo
  {author} {\bibfnamefont {P.}~\bibnamefont {Huillery}}, \bibinfo {author}
  {\bibfnamefont {P.}~\bibnamefont {Pillet}}, \bibinfo {author} {\bibfnamefont
  {D.}~\bibnamefont {Ciampini}}, \bibinfo {author} {\bibfnamefont
  {E.}~\bibnamefont {Arimondo}}, \ and\ \bibinfo {author} {\bibfnamefont
  {O.}~\bibnamefont {Morsch}},\ }\href {\doibase
  10.1103/PhysRevLett.113.023006} {\bibfield  {journal} {\bibinfo  {journal}
  {Phys. Rev. Lett.}\ }\textbf {\bibinfo {volume} {113}},\ \bibinfo {pages}
  {023006} (\bibinfo {year} {2014})}\BibitemShut {NoStop}%
\bibitem [{\citenamefont {Goldschmidt}\ \emph {et~al.}(2016)\citenamefont
  {Goldschmidt}, \citenamefont {Boulier}, \citenamefont {Brown}, \citenamefont
  {Koller}, \citenamefont {Young}, \citenamefont {Gorshkov}, \citenamefont
  {Rolston},\ and\ \citenamefont {Porto}}]{Goldschmidt2016}%
  \BibitemOpen
  \bibfield  {author} {\bibinfo {author} {\bibfnamefont {E.~A.}\ \bibnamefont
  {Goldschmidt}}, \bibinfo {author} {\bibfnamefont {T.}~\bibnamefont
  {Boulier}}, \bibinfo {author} {\bibfnamefont {R.~C.}\ \bibnamefont {Brown}},
  \bibinfo {author} {\bibfnamefont {S.~B.}\ \bibnamefont {Koller}}, \bibinfo
  {author} {\bibfnamefont {J.~T.}\ \bibnamefont {Young}}, \bibinfo {author}
  {\bibfnamefont {A.~V.}\ \bibnamefont {Gorshkov}}, \bibinfo {author}
  {\bibfnamefont {S.~L.}\ \bibnamefont {Rolston}}, \ and\ \bibinfo {author}
  {\bibfnamefont {J.~V.}\ \bibnamefont {Porto}},\ }\href {\doibase
  10.1103/PhysRevLett.116.113001} {\bibfield  {journal} {\bibinfo  {journal}
  {Phys. Rev. Lett.}\ }\textbf {\bibinfo {volume} {116}},\ \bibinfo {pages}
  {113001} (\bibinfo {year} {2016})}\BibitemShut {NoStop}%
\bibitem [{\citenamefont {Helmrich}\ \emph {et~al.}(2018)\citenamefont
  {Helmrich}, \citenamefont {Arias},\ and\ \citenamefont
  {Whitlock}}]{Helmrich2018}%
  \BibitemOpen
  \bibfield  {author} {\bibinfo {author} {\bibfnamefont {S.}~\bibnamefont
  {Helmrich}}, \bibinfo {author} {\bibfnamefont {A.}~\bibnamefont {Arias}}, \
  and\ \bibinfo {author} {\bibfnamefont {S.}~\bibnamefont {Whitlock}},\ }\href
  {\doibase 10.1103/PhysRevA.98.022109} {\bibfield  {journal} {\bibinfo
  {journal} {Phys. Rev. A}\ }\textbf {\bibinfo {volume} {98}},\ \bibinfo
  {pages} {022109} (\bibinfo {year} {2018})}\BibitemShut {NoStop}%
\bibitem [{Note1()}]{Note1}%
  \BibitemOpen
  \bibinfo {note} {The facilitation radius can be approximated as $r_\protect
  \text {fac} = \left (C_6 / \Delta \right )^{1/6}$, where $\Delta $ is the
  detuning and $C_6$ is the van der Waals coefficient.}\BibitemShut {Stop}%
\bibitem [{\citenamefont {Lee}\ \emph {et~al.}(2012)\citenamefont {Lee},
  \citenamefont {H\"affner},\ and\ \citenamefont {Cross}}]{Lee2012}%
  \BibitemOpen
  \bibfield  {author} {\bibinfo {author} {\bibfnamefont {T.~E.}\ \bibnamefont
  {Lee}}, \bibinfo {author} {\bibfnamefont {H.}~\bibnamefont {H\"affner}}, \
  and\ \bibinfo {author} {\bibfnamefont {M.~C.}\ \bibnamefont {Cross}},\ }\href
  {\doibase 10.1103/PhysRevLett.108.023602} {\bibfield  {journal} {\bibinfo
  {journal} {Phys. Rev. Lett.}\ }\textbf {\bibinfo {volume} {108}},\ \bibinfo
  {pages} {023602} (\bibinfo {year} {2012})}\BibitemShut {NoStop}%
\bibitem [{\citenamefont {Lesanovsky}\ and\ \citenamefont
  {Garrahan}(2013)}]{Lesanovsky2013}%
  \BibitemOpen
  \bibfield  {author} {\bibinfo {author} {\bibfnamefont {I.}~\bibnamefont
  {Lesanovsky}}\ and\ \bibinfo {author} {\bibfnamefont {J.~P.}\ \bibnamefont
  {Garrahan}},\ }\href {\doibase 10.1103/PhysRevLett.111.215305} {\bibfield
  {journal} {\bibinfo  {journal} {Phys. Rev. Lett.}\ }\textbf {\bibinfo
  {volume} {111}},\ \bibinfo {pages} {215305} (\bibinfo {year}
  {2013})}\BibitemShut {NoStop}%
\bibitem [{\citenamefont {Marcuzzi}\ \emph {et~al.}(2015)\citenamefont
  {Marcuzzi}, \citenamefont {Levi}, \citenamefont {Li}, \citenamefont
  {Garrahan}, \citenamefont {Olmos},\ and\ \citenamefont
  {Lesanovsky}}]{Marcuzzi2015}%
  \BibitemOpen
  \bibfield  {author} {\bibinfo {author} {\bibfnamefont {M.}~\bibnamefont
  {Marcuzzi}}, \bibinfo {author} {\bibfnamefont {E.}~\bibnamefont {Levi}},
  \bibinfo {author} {\bibfnamefont {W.}~\bibnamefont {Li}}, \bibinfo {author}
  {\bibfnamefont {J.~P.}\ \bibnamefont {Garrahan}}, \bibinfo {author}
  {\bibfnamefont {B.}~\bibnamefont {Olmos}}, \ and\ \bibinfo {author}
  {\bibfnamefont {I.}~\bibnamefont {Lesanovsky}},\ }\href {\doibase
  10.1088/1367-2630/17/7/072003} {\bibfield  {journal} {\bibinfo  {journal}
  {New Journal of Physics}\ }\textbf {\bibinfo {volume} {17}},\ \bibinfo
  {pages} {072003} (\bibinfo {year} {2015})},\ \Eprint
  {http://arxiv.org/abs/1411.7984} {1411.7984} \BibitemShut {NoStop}%
\bibitem [{\citenamefont {Marcuzzi}\ \emph {et~al.}(2016)\citenamefont
  {Marcuzzi}, \citenamefont {Buchhold}, \citenamefont {Diehl},\ and\
  \citenamefont {Lesanovsky}}]{Marcuzzi2016}%
  \BibitemOpen
  \bibfield  {author} {\bibinfo {author} {\bibfnamefont {M.}~\bibnamefont
  {Marcuzzi}}, \bibinfo {author} {\bibfnamefont {M.}~\bibnamefont {Buchhold}},
  \bibinfo {author} {\bibfnamefont {S.}~\bibnamefont {Diehl}}, \ and\ \bibinfo
  {author} {\bibfnamefont {I.}~\bibnamefont {Lesanovsky}},\ }\href {\doibase
  10.1103/PhysRevLett.116.245701} {\bibfield  {journal} {\bibinfo  {journal}
  {Phys. Rev. Lett.}\ }\textbf {\bibinfo {volume} {116}},\ \bibinfo {pages}
  {245701} (\bibinfo {year} {2016})}\BibitemShut {NoStop}%
\bibitem [{\citenamefont {L\"ow}\ \emph {et~al.}(2012)\citenamefont {L\"ow},
  \citenamefont {Weimer}, \citenamefont {Nipper}, \citenamefont {Balewski},
  \citenamefont {Butscher}, \citenamefont {B\"uchler},\ and\ \citenamefont
  {Pfau}}]{Weimer2012}%
  \BibitemOpen
  \bibfield  {author} {\bibinfo {author} {\bibfnamefont {R.}~\bibnamefont
  {L\"ow}}, \bibinfo {author} {\bibfnamefont {H.}~\bibnamefont {Weimer}},
  \bibinfo {author} {\bibfnamefont {J.}~\bibnamefont {Nipper}}, \bibinfo
  {author} {\bibfnamefont {J.~B.}\ \bibnamefont {Balewski}}, \bibinfo {author}
  {\bibfnamefont {B.}~\bibnamefont {Butscher}}, \bibinfo {author}
  {\bibfnamefont {H.~P.}\ \bibnamefont {B\"uchler}}, \ and\ \bibinfo {author}
  {\bibfnamefont {T.}~\bibnamefont {Pfau}},\ }\href
  {http://stacks.iop.org/0953-4075/45/i=11/a=113001} {\bibfield  {journal}
  {\bibinfo  {journal} {Journal of Physics B: Atomic, Molecular and Optical
  Physics}\ }\textbf {\bibinfo {volume} {45}},\ \bibinfo {pages} {113001}
  (\bibinfo {year} {2012})}\BibitemShut {NoStop}%
\bibitem [{\citenamefont {Heidemann}\ \emph {et~al.}(2007)\citenamefont
  {Heidemann}, \citenamefont {Raitzsch}, \citenamefont {Bendkowsky},
  \citenamefont {Butscher}, \citenamefont {L{\"{o}}w}, \citenamefont {Santos},\
  and\ \citenamefont {Pfau}}]{Heidemann2007}%
  \BibitemOpen
  \bibfield  {author} {\bibinfo {author} {\bibfnamefont {R.}~\bibnamefont
  {Heidemann}}, \bibinfo {author} {\bibfnamefont {U.}~\bibnamefont {Raitzsch}},
  \bibinfo {author} {\bibfnamefont {V.}~\bibnamefont {Bendkowsky}}, \bibinfo
  {author} {\bibfnamefont {B.}~\bibnamefont {Butscher}}, \bibinfo {author}
  {\bibfnamefont {R.}~\bibnamefont {L{\"{o}}w}}, \bibinfo {author}
  {\bibfnamefont {L.}~\bibnamefont {Santos}}, \ and\ \bibinfo {author}
  {\bibfnamefont {T.}~\bibnamefont {Pfau}},\ }\href@noop {} {\bibfield
  {journal} {\bibinfo  {journal} {Phys. Rev. Lett.}\ }\textbf {\bibinfo
  {volume} {99}},\ \bibinfo {pages} {163601} (\bibinfo {year}
  {2007})}\BibitemShut {NoStop}%
\bibitem [{\citenamefont {Urvoy}\ \emph {et~al.}(2015)\citenamefont {Urvoy},
  \citenamefont {Ripka}, \citenamefont {Lesanovsky}, \citenamefont {Booth},
  \citenamefont {Shaffer}, \citenamefont {Pfau},\ and\ \citenamefont
  {L\"ow}}]{Urvoy2015a}%
  \BibitemOpen
  \bibfield  {author} {\bibinfo {author} {\bibfnamefont {A.}~\bibnamefont
  {Urvoy}}, \bibinfo {author} {\bibfnamefont {F.}~\bibnamefont {Ripka}},
  \bibinfo {author} {\bibfnamefont {I.}~\bibnamefont {Lesanovsky}}, \bibinfo
  {author} {\bibfnamefont {D.}~\bibnamefont {Booth}}, \bibinfo {author}
  {\bibfnamefont {J.~P.}\ \bibnamefont {Shaffer}}, \bibinfo {author}
  {\bibfnamefont {T.}~\bibnamefont {Pfau}}, \ and\ \bibinfo {author}
  {\bibfnamefont {R.}~\bibnamefont {L\"ow}},\ }\href {\doibase
  10.1103/PhysRevLett.114.203002} {\bibfield  {journal} {\bibinfo  {journal}
  {Phys. Rev. Lett.}\ }\textbf {\bibinfo {volume} {114}},\ \bibinfo {pages}
  {203002} (\bibinfo {year} {2015})}\BibitemShut {NoStop}%
\bibitem [{\citenamefont {Weimer}\ \emph {et~al.}(2008)\citenamefont {Weimer},
  \citenamefont {L{\"{o}}w}, \citenamefont {Pfau},\ and\ \citenamefont
  {B{\"{u}}chler}}]{Weimer2008}%
  \BibitemOpen
  \bibfield  {author} {\bibinfo {author} {\bibfnamefont {H.}~\bibnamefont
  {Weimer}}, \bibinfo {author} {\bibfnamefont {R.}~\bibnamefont {L{\"{o}}w}},
  \bibinfo {author} {\bibfnamefont {T.}~\bibnamefont {Pfau}}, \ and\ \bibinfo
  {author} {\bibfnamefont {H.~P.}\ \bibnamefont {B{\"{u}}chler}},\ }\href
  {\doibase 10.1103/PhysRevLett.101.250601} {\bibfield  {journal} {\bibinfo
  {journal} {Phys. Rev. Lett.}\ }\textbf {\bibinfo {volume} {101}},\ \bibinfo
  {pages} {250601} (\bibinfo {year} {2008})}\BibitemShut {NoStop}%
\bibitem [{\citenamefont {Ates}\ \emph {et~al.}(2007)\citenamefont {Ates},
  \citenamefont {Pohl}, \citenamefont {Pattard},\ and\ \citenamefont
  {Rost}}]{Ates2007a}%
  \BibitemOpen
  \bibfield  {author} {\bibinfo {author} {\bibfnamefont {C.}~\bibnamefont
  {Ates}}, \bibinfo {author} {\bibfnamefont {T.}~\bibnamefont {Pohl}}, \bibinfo
  {author} {\bibfnamefont {T.}~\bibnamefont {Pattard}}, \ and\ \bibinfo
  {author} {\bibfnamefont {J.~M.}\ \bibnamefont {Rost}},\ }\href {\doibase
  10.1103/PhysRevLett.98.023002} {\bibfield  {journal} {\bibinfo  {journal}
  {Phys. Rev. Lett.}\ }\textbf {\bibinfo {volume} {98}},\ \bibinfo {pages}
  {023002} (\bibinfo {year} {2007})}\BibitemShut {NoStop}%
\bibitem [{\citenamefont {Amthor}\ \emph {et~al.}(2010)\citenamefont {Amthor},
  \citenamefont {Giese}, \citenamefont {Hofmann},\ and\ \citenamefont
  {Weidem\"uller}}]{Amthor2010}%
  \BibitemOpen
  \bibfield  {author} {\bibinfo {author} {\bibfnamefont {T.}~\bibnamefont
  {Amthor}}, \bibinfo {author} {\bibfnamefont {C.}~\bibnamefont {Giese}},
  \bibinfo {author} {\bibfnamefont {C.~S.}\ \bibnamefont {Hofmann}}, \ and\
  \bibinfo {author} {\bibfnamefont {M.}~\bibnamefont {Weidem\"uller}},\ }\href
  {\doibase 10.1103/PhysRevLett.104.013001} {\bibfield  {journal} {\bibinfo
  {journal} {Phys. Rev. Lett.}\ }\textbf {\bibinfo {volume} {104}},\ \bibinfo
  {pages} {013001} (\bibinfo {year} {2010})}\BibitemShut {NoStop}%
\bibitem [{\citenamefont {G\"arttner}\ \emph {et~al.}(2013)\citenamefont
  {G\"arttner}, \citenamefont {Heeg}, \citenamefont {Gasenzer},\ and\
  \citenamefont {Evers}}]{Gart2013}%
  \BibitemOpen
  \bibfield  {author} {\bibinfo {author} {\bibfnamefont {M.}~\bibnamefont
  {G\"arttner}}, \bibinfo {author} {\bibfnamefont {K.~P.}\ \bibnamefont
  {Heeg}}, \bibinfo {author} {\bibfnamefont {T.}~\bibnamefont {Gasenzer}}, \
  and\ \bibinfo {author} {\bibfnamefont {J.}~\bibnamefont {Evers}},\ }\href
  {\doibase 10.1103/PhysRevA.88.043410} {\bibfield  {journal} {\bibinfo
  {journal} {Phys. Rev. A}\ }\textbf {\bibinfo {volume} {88}},\ \bibinfo
  {pages} {043410} (\bibinfo {year} {2013})}\BibitemShut {NoStop}%
\bibitem [{\citenamefont {Simonelli}\ \emph {et~al.}(2016)\citenamefont
  {Simonelli}, \citenamefont {Valado}, \citenamefont {Masella}, \citenamefont
  {Asteria}, \citenamefont {Arimondo}, \citenamefont {Ciampini},\ and\
  \citenamefont {Morsch}}]{Simonelli2016}%
  \BibitemOpen
  \bibfield  {author} {\bibinfo {author} {\bibfnamefont {C.}~\bibnamefont
  {Simonelli}}, \bibinfo {author} {\bibfnamefont {M.~M.}\ \bibnamefont
  {Valado}}, \bibinfo {author} {\bibfnamefont {G.}~\bibnamefont {Masella}},
  \bibinfo {author} {\bibfnamefont {L.}~\bibnamefont {Asteria}}, \bibinfo
  {author} {\bibfnamefont {E.}~\bibnamefont {Arimondo}}, \bibinfo {author}
  {\bibfnamefont {D.}~\bibnamefont {Ciampini}}, \ and\ \bibinfo {author}
  {\bibfnamefont {O.}~\bibnamefont {Morsch}},\ }\href
  {http://stacks.iop.org/0953-4075/49/i=15/a=154002} {\bibfield  {journal}
  {\bibinfo  {journal} {Journal of Physics B: Atomic, Molecular and Optical
  Physics}\ }\textbf {\bibinfo {volume} {49}},\ \bibinfo {pages} {154002}
  (\bibinfo {year} {2016})}\BibitemShut {NoStop}%
\bibitem [{\citenamefont {Faoro}\ \emph {et~al.}(2016)\citenamefont {Faoro},
  \citenamefont {Simonelli}, \citenamefont {Archimi}, \citenamefont {Masella},
  \citenamefont {Valado}, \citenamefont {Arimondo}, \citenamefont {Mannella},
  \citenamefont {Ciampini},\ and\ \citenamefont {Morsch}}]{Morsch2016a}%
  \BibitemOpen
  \bibfield  {author} {\bibinfo {author} {\bibfnamefont {R.}~\bibnamefont
  {Faoro}}, \bibinfo {author} {\bibfnamefont {C.}~\bibnamefont {Simonelli}},
  \bibinfo {author} {\bibfnamefont {M.}~\bibnamefont {Archimi}}, \bibinfo
  {author} {\bibfnamefont {G.}~\bibnamefont {Masella}}, \bibinfo {author}
  {\bibfnamefont {M.~M.}\ \bibnamefont {Valado}}, \bibinfo {author}
  {\bibfnamefont {E.}~\bibnamefont {Arimondo}}, \bibinfo {author}
  {\bibfnamefont {R.}~\bibnamefont {Mannella}}, \bibinfo {author}
  {\bibfnamefont {D.}~\bibnamefont {Ciampini}}, \ and\ \bibinfo {author}
  {\bibfnamefont {O.}~\bibnamefont {Morsch}},\ }\href {\doibase
  10.1103/PhysRevA.93.030701} {\bibfield  {journal} {\bibinfo  {journal} {Phys.
  Rev. A}\ }\textbf {\bibinfo {volume} {93}},\ \bibinfo {pages} {030701}
  (\bibinfo {year} {2016})}\BibitemShut {NoStop}%
\bibitem [{\citenamefont {Lesanovsky}\ and\ \citenamefont
  {Garrahan}(2014)}]{Lesa2014}%
  \BibitemOpen
  \bibfield  {author} {\bibinfo {author} {\bibfnamefont {I.}~\bibnamefont
  {Lesanovsky}}\ and\ \bibinfo {author} {\bibfnamefont {J.~P.}\ \bibnamefont
  {Garrahan}},\ }\href {\doibase 10.1103/PhysRevA.90.011603} {\bibfield
  {journal} {\bibinfo  {journal} {Phys. Rev. A}\ }\textbf {\bibinfo {volume}
  {90}},\ \bibinfo {pages} {011603} (\bibinfo {year} {2014})}\BibitemShut
  {NoStop}%
\bibitem [{\citenamefont {Valado}\ \emph {et~al.}(2016)\citenamefont {Valado},
  \citenamefont {Simonelli}, \citenamefont {Hoogerland}, \citenamefont
  {Lesanovsky}, \citenamefont {Garrahan}, \citenamefont {Arimondo},
  \citenamefont {Ciampini},\ and\ \citenamefont {Morsch}}]{Morsch2016}%
  \BibitemOpen
  \bibfield  {author} {\bibinfo {author} {\bibfnamefont {M.~M.}\ \bibnamefont
  {Valado}}, \bibinfo {author} {\bibfnamefont {C.}~\bibnamefont {Simonelli}},
  \bibinfo {author} {\bibfnamefont {M.~D.}\ \bibnamefont {Hoogerland}},
  \bibinfo {author} {\bibfnamefont {I.}~\bibnamefont {Lesanovsky}}, \bibinfo
  {author} {\bibfnamefont {J.~P.}\ \bibnamefont {Garrahan}}, \bibinfo {author}
  {\bibfnamefont {E.}~\bibnamefont {Arimondo}}, \bibinfo {author}
  {\bibfnamefont {D.}~\bibnamefont {Ciampini}}, \ and\ \bibinfo {author}
  {\bibfnamefont {O.}~\bibnamefont {Morsch}},\ }\href {\doibase
  10.1103/PhysRevA.93.040701} {\bibfield  {journal} {\bibinfo  {journal} {Phys.
  Rev. A}\ }\textbf {\bibinfo {volume} {93}},\ \bibinfo {pages} {040701}
  (\bibinfo {year} {2016})}\BibitemShut {NoStop}%
\bibitem [{\citenamefont {Pupillo}\ \emph {et~al.}(2010)\citenamefont
  {Pupillo}, \citenamefont {Micheli}, \citenamefont {Boninsegni}, \citenamefont
  {Lesanovsky},\ and\ \citenamefont {Zoller}}]{Pupillo2010}%
  \BibitemOpen
  \bibfield  {author} {\bibinfo {author} {\bibfnamefont {G.}~\bibnamefont
  {Pupillo}}, \bibinfo {author} {\bibfnamefont {A.}~\bibnamefont {Micheli}},
  \bibinfo {author} {\bibfnamefont {M.}~\bibnamefont {Boninsegni}}, \bibinfo
  {author} {\bibfnamefont {I.}~\bibnamefont {Lesanovsky}}, \ and\ \bibinfo
  {author} {\bibfnamefont {P.}~\bibnamefont {Zoller}},\ }\href {\doibase
  10.1103/PhysRevLett.104.223002} {\bibfield  {journal} {\bibinfo  {journal}
  {Phys. Rev. Lett.}\ }\textbf {\bibinfo {volume} {104}},\ \bibinfo {pages}
  {223002} (\bibinfo {year} {2010})}\BibitemShut {NoStop}%
\bibitem [{\citenamefont {Hickstein}\ \emph {et~al.}(2019)\citenamefont
  {Hickstein}, \citenamefont {Gibson}, \citenamefont {Yurchak}, \citenamefont
  {Das},\ and\ \citenamefont {Ryazanov}}]{hickstein2019PyAbel}%
  \BibitemOpen
  \bibfield  {author} {\bibinfo {author} {\bibfnamefont {D.~D.}\ \bibnamefont
  {Hickstein}}, \bibinfo {author} {\bibfnamefont {S.~T.}\ \bibnamefont
  {Gibson}}, \bibinfo {author} {\bibfnamefont {R.}~\bibnamefont {Yurchak}},
  \bibinfo {author} {\bibfnamefont {D.~D.}\ \bibnamefont {Das}}, \ and\
  \bibinfo {author} {\bibfnamefont {M.}~\bibnamefont {Ryazanov}},\ }\href
  {\doibase 10.1063/1.5092635} {\bibfield  {journal} {\bibinfo  {journal}
  {Review of Scientific Instruments}\ }\textbf {\bibinfo {volume} {90}},\
  \bibinfo {pages} {065115} (\bibinfo {year} {2019})}\BibitemShut {NoStop}%
\bibitem [{\citenamefont {Buchhold}\ \emph {et~al.}(2017)\citenamefont
  {Buchhold}, \citenamefont {Everest}, \citenamefont {Marcuzzi}, \citenamefont
  {Lesanovsky},\ and\ \citenamefont {Diehl}}]{Buchhold2017}%
  \BibitemOpen
  \bibfield  {author} {\bibinfo {author} {\bibfnamefont {M.}~\bibnamefont
  {Buchhold}}, \bibinfo {author} {\bibfnamefont {B.}~\bibnamefont {Everest}},
  \bibinfo {author} {\bibfnamefont {M.}~\bibnamefont {Marcuzzi}}, \bibinfo
  {author} {\bibfnamefont {I.}~\bibnamefont {Lesanovsky}}, \ and\ \bibinfo
  {author} {\bibfnamefont {S.}~\bibnamefont {Diehl}},\ }\href {\doibase
  10.1103/PhysRevB.95.014308} {\bibfield  {journal} {\bibinfo  {journal} {Phys.
  Rev. B}\ }\textbf {\bibinfo {volume} {95}},\ \bibinfo {pages} {014308}
  (\bibinfo {year} {2017})}\BibitemShut {NoStop}%
\bibitem [{sup()}]{supp_noto4}%
  \BibitemOpen
  \href@noop {} {}\bibinfo {note} {{See supplementary material appended to this
  manuscript.}}\BibitemShut {Stop}%
\bibitem [{\citenamefont {Dornic}\ \emph {et~al.}(2005)\citenamefont {Dornic},
  \citenamefont {Chat\'e},\ and\ \citenamefont {Mu\~noz}}]{Dornic}%
  \BibitemOpen
  \bibfield  {author} {\bibinfo {author} {\bibfnamefont {I.}~\bibnamefont
  {Dornic}}, \bibinfo {author} {\bibfnamefont {H.}~\bibnamefont {Chat\'e}}, \
  and\ \bibinfo {author} {\bibfnamefont {M.~A.}\ \bibnamefont {Mu\~noz}},\
  }\href {\doibase 10.1103/PhysRevLett.94.100601} {\bibfield  {journal}
  {\bibinfo  {journal} {Phys. Rev. Lett.}\ }\textbf {\bibinfo {volume} {94}},\
  \bibinfo {pages} {100601} (\bibinfo {year} {2005})}\BibitemShut {NoStop}%
\bibitem [{\citenamefont {Klocke}\ and\ \citenamefont
  {Buchhold}(2019)}]{Klocke}%
  \BibitemOpen
  \bibfield  {author} {\bibinfo {author} {\bibfnamefont {K.}~\bibnamefont
  {Klocke}}\ and\ \bibinfo {author} {\bibfnamefont {M.}~\bibnamefont
  {Buchhold}},\ }\href {\doibase 10.1103/physreva.99.053616} {\bibfield
  {journal} {\bibinfo  {journal} {Phys. Rev. A}\ }\textbf {\bibinfo {volume}
  {99}},\ \bibinfo {pages} {053616} (\bibinfo {year} {2019})}\BibitemShut
  {NoStop}%
\bibitem [{not()}]{noto3}%
  \BibitemOpen
  \href@noop {} {}\bibinfo {note} {{For simulations we take a 30 $\times$ 30
  $\times$ 600 spatial grid and a time discretization $\Delta t = 0.002$ with
  $\kappa=5.2$, $\Gamma = 5.3$, $\gamma_{\downarrow 0} = 1.5$, $D = 10$, $\tau
  = 6\times10^{-7}$, $D_T = 9$
  }}\BibitemShut {NoStop}%
\bibitem [{\citenamefont {Wintermantel}\ \emph {et~al.}(2020)\citenamefont
  {Wintermantel}, \citenamefont {Buchhold}, \citenamefont {Shevate},
  \citenamefont {Morgado}, \citenamefont {Wang}, \citenamefont {Lochead},
  \citenamefont {Diehl},\ and\ \citenamefont {Whitlock}}]{Wintermantel2020}%
  \BibitemOpen
  \bibfield  {author} {\bibinfo {author} {\bibfnamefont {T.~M.}\ \bibnamefont
  {Wintermantel}}, \bibinfo {author} {\bibfnamefont {M.}~\bibnamefont
  {Buchhold}}, \bibinfo {author} {\bibfnamefont {S.}~\bibnamefont {Shevate}},
  \bibinfo {author} {\bibfnamefont {M.}~\bibnamefont {Morgado}}, \bibinfo
  {author} {\bibfnamefont {Y.}~\bibnamefont {Wang}}, \bibinfo {author}
  {\bibfnamefont {G.}~\bibnamefont {Lochead}}, \bibinfo {author} {\bibfnamefont
  {S.}~\bibnamefont {Diehl}}, \ and\ \bibinfo {author} {\bibfnamefont
  {S.}~\bibnamefont {Whitlock}},\ }\href@noop {} {\enquote {\bibinfo {title}
  {Epidemic growth and griffiths effects on an emergent network of excited
  atoms},}\ } (\bibinfo {year} {2020}),\ \Eprint
  {http://arxiv.org/abs/2007.07697} {arXiv:2007.07697} \BibitemShut {NoStop}%
\bibitem [{\citenamefont {Wang}\ \emph {et~al.}(2020)\citenamefont {Wang},
  \citenamefont {Aghigh}, \citenamefont {Marroquín}, \citenamefont {Grant},
  \citenamefont {Sous}, \citenamefont {Martins}, \citenamefont {Keller},\ and\
  \citenamefont {Grant}}]{wang2020}%
  \BibitemOpen
  \bibfield  {author} {\bibinfo {author} {\bibfnamefont {R.}~\bibnamefont
  {Wang}}, \bibinfo {author} {\bibfnamefont {M.}~\bibnamefont {Aghigh}},
  \bibinfo {author} {\bibfnamefont {K.~L.}\ \bibnamefont {Marroquín}},
  \bibinfo {author} {\bibfnamefont {K.~M.}\ \bibnamefont {Grant}}, \bibinfo
  {author} {\bibfnamefont {J.}~\bibnamefont {Sous}}, \bibinfo {author}
  {\bibfnamefont {F.~B.~V.}\ \bibnamefont {Martins}}, \bibinfo {author}
  {\bibfnamefont {J.~S.}\ \bibnamefont {Keller}}, \ and\ \bibinfo {author}
  {\bibfnamefont {E.~R.}\ \bibnamefont {Grant}},\ }\href@noop {} {} (\bibinfo
  {year} {2020}),\ \Eprint {http://arxiv.org/abs/2006.16412} {arXiv:2006.16412}
  \BibitemShut {NoStop}%
\bibitem [{\citenamefont {Buend\'ia}\ \emph {et~al.}(2020)\citenamefont
  {Buend\'ia}, \citenamefont {di~Santo}, \citenamefont {Bonachela},\ and\
  \citenamefont {Mu\~noz}}]{buenda2020feedback}%
  \BibitemOpen
  \bibfield  {author} {\bibinfo {author} {\bibfnamefont {V.}~\bibnamefont
  {Buend\'ia}}, \bibinfo {author} {\bibfnamefont {S.}~\bibnamefont {di~Santo}},
  \bibinfo {author} {\bibfnamefont {J.~A.}\ \bibnamefont {Bonachela}}, \ and\
  \bibinfo {author} {\bibfnamefont {M.~A.}\ \bibnamefont {Mu\~noz}},\
  }\href@noop {} {\  (\bibinfo {year} {2020})},\ \Eprint
  {http://arxiv.org/abs/2006.03020} {arXiv:2006.03020} \BibitemShut {NoStop}%
\end{thebibliography}%

\clearpage
\setcounter{secnumdepth}{2} 
\appendix

\begin{widetext}

\section{Numerical integration scheme}

We simulate the Langevin equations on a discrete spatio-temporal lattice by means of an operator-splitting update scheme. Consider starting from a general stochastic differential equation (SDE) over the density field $\rho_{\vec{x},t}$ like

\begin{equation}
    \partial_t \rho_{\vec{x},t} = D\nabla^2\rho_{\vec{x},t} + a + b\rho_{\vec{x,t}} + c\rho^2_{\vec{x},t} + \sigma^2\sqrt{\rho_{\vec{x},t} + d}\eta,\label{eq:genericSDE}
\end{equation}

where the Markovian noise kernel $\eta$ has unit variance and zero mean. On a lattice the Laplacian is discretized so that it gets absorbed into the coefficients $a$ and $b$. Then under appropriate change of variables $\rho \rightarrow \rho' \equiv \rho + d$ we may eliminate the constant offset in the noise term. The temporal update is then decomposed into two steps: a stochastic step and a deterministic step. For the former we drop the quadratic term, yielding an SDE of the form

\begin{equation}
    \partial_t \rho_{\vec{x},t} = \alpha + \beta\rho_{\vec{x},t} + \sigma\sqrt{\rho}\eta.
\end{equation}

This class of SDEs with a multiplicative noise kernel admits an exact solution to the corresponding Fokker-Planck equation. Denoting the current value as $\rho_0$, the distribution of values $\rho$ after a time step $\delta t$ is given by

\begin{equation}
    P(\rho) = \lambda e^{-\lambda\left(\rho_0e^{\beta\delta t} + \rho\right)}\left(\frac{\rho}{\rho_0e^{\beta\delta t}}\right)^{\mu/2}I_\mu \left(2\lambda\sqrt{\rho_0\rho e^{\beta \delta t}}\right).
\end{equation}

Here we have denoted $\lambda = \frac{2\beta}{\sigma^2(e^{\beta \delta t} - 1)}$ and $\mu = \frac{2\alpha}{\sigma^2} - 1$. This distribution may be efficiently sampled by rewriting it as a mixed Gamma distribution 

\begin{equation}
    \rho = \Gamma[\mu + 1 + \text{Poisson}[\lambda\rho_0 e^{\beta\delta t}]]/\lambda.
\end{equation}

The stochastic evolution of the state at time $t$ reduces to randomly sampling from the above distribution at each time step. For $d\not= 0$ in Eq.~\eqref{eq:genericSDE} this requires that we enforce non-negativity of $\rho$ in terms of the transformed variables by taking all $\rho' < d$ to $d$. The remaining deterministic quadratic term may be treated by a standard Euler scheme. Similarly, we treat the evolution of the total density $n_{\vec{x},t}$ by an Euler scheme.

\section{Equations of motion}
The equations of motion for the density $n_{\vec{x},t}, \rho_{\vec{x},t}$ follow from acting the derivative on their definition and applying the chain rule
\begin{equation}\label{EOM1_supp}
\partial_tn_{\vec{x},t}=\partial_t \sum_l \langle \sigma^{rr}_l+\sigma^{gg}_l\rangle \theta(r_{\text{fac}}^2-|\vec{r}_l-\vec{x}|^2)=\sum_l \theta(r_{\text{fac}}^2-|\vec{r}_l-\vec{x}|^2) \langle \partial_t\sigma^{rr}_l+\partial_t\sigma^{gg}_l\rangle+\sum_l \langle \sigma^{rr}_l+\sigma^{gg}_l\rangle\nabla_{\vec{r}_l}\theta(r_{\text{fac}}^2-|\vec{r}_l-\vec{x}|^2)\partial_t\vec{r}_l.
\end{equation}
The first part on the RHS of Eq.~\eqref{EOM1_supp} describes the evolution of the internal degrees of freedom of the particles in the unit cell. It has been discussed in detail in Ref.~\cite{Buchhold2017} and it captures the loss of particles into the dark state. The second part considers the evolution of the motional degrees of freedom. Evaluating the equations of motion $\partial_t\langle\hat{O}\rangle=\text{Tr}(\hat{O}\partial_t\rho)$ with the master equation from the main text yields
\begin{equation}
    \partial_tn_{\vec{x},t}=-\gamma_{\downarrow0}\rho_{\vec{x},t}-\nabla_{\vec{x}} \underbrace{\sum_l \langle \sigma^{rr}_l+\sigma^{gg}_l\rangle\theta(r_{\text{fac}}^2-|\vec{r}_l-\vec{x}|^2)\vec{v}_l}_{\vec{j}_{\vec{x}}}.
\end{equation}
Here, we have replaced $\vec{v}_l=\partial_t\vec{r}_l$ and we have exploited $\nabla_{\vec{r}_l}f(\vec{r}_l-\vec{x})=-\nabla_{\vec{x}}f(\vec{r}_l-\vec{x})$ in order to pull the gradient out of the sum. The average over all the particles' velocities is the coarse grained current $\vec{j}$ in the unit cell. This yields the equation of motion for $n_{\vec{x},t}$ in the main text. The current can be evaluated using the classical equations of motion in the relaxation time approximation. The Hamilton equation of motion for the velocity in the presence of a potential $V$ and subject to random collisions (i.e. Brownian motion) reads as
\begin{equation}
    M\partial_t\vec{v}_l=-\nabla_{\vec{r}_l}V(\vec{r}_l)-\frac{1}{\eta} \vec{v}_l + \vec{\xi}_{t,\vec{r}_l}
\end{equation}
where $\eta=\frac{d_a}{\sqrt{Mk_BT}}$ is the mobility. In an ideal gas it is set by the ratio of the mean free path $d_a$ of an atom and the deBroglie wavelength $\lambda_{db}=\sqrt{Mk_BT}$. The random forces $\vec{\xi}_{t,\vec{r}_l}$ describe collisions with other atoms. In the overdamped case $\vec{v}_l=-\eta\nabla_{\vec{r}_l}V(\vec{r}_l)+\eta\vec{\xi}_{t,\vec{r}_l}$. One can insert this result into the definition of the current
\begin{equation}
    \vec{j}_{\vec{x}}=\sum_l \langle \sigma^{rr}_l+\sigma^{gg}_l\rangle\theta(r_{\text{fac}}^2-|\vec{r}_l-\vec{x}|^2)(-\eta\nabla_{\vec{r}_l}V(\vec{r}_l)+\eta\vec{\xi}_{t,\vec{r}_l})=-\eta n_{\vec{x},t}\nabla_{\vec{x}}V(\vec{x})-D_T\nabla n_{\vec{x},t}.
\end{equation}
Here, we first approximated $V(\vec{r}_l)$ by $V(\vec{x})$ for all $\vec{r}_l$ in the unit cell in order to pull the average potential force out of the sum. This is justified as long as the potential varies on much larger scales than $r_{\text{fac}}\ll w$. Second, we made the intuitive approximation that the average over the random forces felt by an atom in the unit cell (which originate from collisions with the other atoms) pushes the atom towards the region of the lowest instantaneous density, which is given by $-\nabla n_{\vec{x},t}$. This is the common relaxation time approximation, leading to Brownian diffusion of the density. The proportionality constant $D_T=\eta k_B T$ follows from the Einstein relation. 

\end{widetext}

\end{document}